\documentclass[aps,prb,twocolumn,superscriptaddress]{revtex4-2}

\usepackage{amsmath , amssymb}
\usepackage{mathrsfs}
\usepackage[dvipdfmx]{graphicx}
\usepackage{bm}
\usepackage{braket}
\usepackage{color}

\newcommand{\diff}{\mathrm{d}}
\newcommand{\imag}{\mathrm{Im}\,}
\newcommand{\Hc}{\mathrm{H.c.}}
\newcommand{\real}{\mathrm{Re}\,}
\newcommand{\trace}{\mathrm{Tr}\,}
\newcommand{\imu}{\mathrm{i}}
\newcommand{\epn}{\mathrm{e}}
\newcommand{\sgn}{\mathrm{sgn}\,}
\newcommand{\ua}{\uparrow}
\newcommand{\da}{\downarrow}
\newcommand{\dg}{\dagger}
\newcommand{\la}{\langle}
\newcommand{\ra}{\rangle}
\newcommand{\al}{\alpha}
\newcommand{\be}{\beta}
\newcommand{\sg}{\sigma}
\newcommand{\om}{\omega}
\newcommand{\gm}{\gamma}
\newcommand{\lam}{\lambda}
\newcommand{\ep}{\varepsilon}
\newcommand{\T}{\mathrm{T}}
\newcommand{\nt}{\notag \\}

\newcommand{\mrm}[1]{\mathrm{#1}}
\newcommand{\mcal}[1]{\mathcal{#1}}
\newcommand{\mscr}[1]{\mathscr{#1}}

\begin{document}

\title{
Impurity effect on Bogoliubov Fermi surfaces: \\
Analysis based on iron-based superconductors
}

\author{Tatsuya Miki}
\affiliation{Department of Physics, Saitama University, Shimo-Okubo, Saitama 338-8570, Japan}
\author{Hiroaki Ikeda}
\affiliation{Department of Physics, Ritsumeikan University, Kusatsu 525-8577, Japan}
\author{Shintaro Hoshino}
\affiliation{Department of Physics, Saitama University, Shimo-Okubo, Saitama 338-8570, Japan}

\date{\today}

\begin{abstract}
The effect of impurities on a superconductor with Bogoliubov Fermi surfaces (BFSs) is studied using a realistic tight-binding model. Based on the band structure composed of $d$-orbitals in tetragonal $\mrm{FeSe}$, whose S-doped sample is a potential material for BFS, we construct the superconducting state by introducing a time-reversal broken pair potential in terms of the band index. We further consider the effect of impurities on the BFS, where the impurity potential is defined as a local potential for the original $d$-orbitals. The self-energy is calculated using the (self-consistent) Born approximation, which shows an enhancement of the single-particle spectral weight on the Fermi surface. This is consistent with the previous phenomenological theory and is justified by the present more detailed calculation based on the $\mrm{FeSe}$-based material. 
\end{abstract}

\maketitle

\section{Introduction}

The phenomenon of superconductivity is induced by Cooper-pair condensation near the Fermi surface, which typically results in the formation of a superconducting gap at the Fermi level.
This gap structure is usually classified into three categories: full-gap, point-node, and line-node \cite{Sigrist91}.
However, it has been suggested that some superconductors exhibit a fourth type of gap structure known as Bogoliubov Fermi surface (BFS) or ultra-nodal pair, in which the Fermi surface persists even in the superconducting state.
This type of superconductor was first proposed in the context of multi-band superconductors \cite{Volovik89-1, Volovik89-2}, superfluid helium \cite{Liu03, Gubankova05, Autti20}.
More recently, specific models with broken time-reversal symmetry and preserved inversion symmetry have also been proposed as candidate systems \cite{Agterberg17, Brydon18}.
In these models, BFS is topologically protected and remains stable against small perturbations.
The characteristic features of BFS have been the subject of theoretical studies \cite{
Volovik93, Yuan18, Sumita19, Suh20, Setty20, Setty20-prb, Lapp20, Oh20, Tamura20,  Timm21, Timm21_2, Jiang21, Miki21, Hoshino22, Menke19, Link20, Link20-2, Herbut20, Dutta21, Kim21, Kobayashi22, Kitamura22}.

In addition to these theoretical studies, the possibility of BFS has been experimentally implied.
It has been pointed out in some materials with unconventional superconductivity that there exists a residual zero-energy density of state (DOS) in the superconducting state \cite{Schuberth92, Zieve04, Kittaka18, Shibauchi20}. 
Especially in $\mrm{Fe(Se,S)}$, the zero-energy DOS and the presence of low-energy carriers have been observed through the tunnel conductance of STS \cite{Hanaguri18}, the heat capacity, the thermal conductivity \cite{Sato18, Mizukami21}, and the laser ARPES \cite{Nagashima22}.
As for the theoretical description of BFS in $\mrm{Fe(Se,S)}$, the inter-band pairing with broken time reversal symmetry is suggested to play an important role for the system having the BFS \cite{Setty20}.
This model has succeeded in qualitatively reproducing the behavior of the DOS and the heat capacity.

Given the fact that the actual materials may have BFS, it is interesting to ask if there exists characteristic physics specific to superconductors with BFS.
The electronic states near BFS is composed of Bogoliubov quasi-particles (bogolons), which describe the low energy excitation of the superconducting state. 
It is expected that the low-energy properties are governed by the bogolon's nature.
In our previous work, the authors pointed out that the characteristic feature of bogolon enters through the impurity scattering and interaction \cite{Miki21}.
We studied physical properties of the Bogoliubov Fermi liquid state near BFS for a system with preserved inversion and broken time-reversal symmetries, and found that the pair amplitude (anomalous Green's function) of bogolons becomes finite.
Interestingly, the pair amplitude has a purely odd-function with respect to the relative time of two bogolons, which is called the odd-frequency pairing.
The concept of odd-frequency pairing has been previously examined in relation to electrons and $^3$He \cite{Berezinskii74, Kirkpatrick91, Balatsky92, Emery92, Coleman93, Tanaka12, Linder19}, but in the present context the Cooper pair is composed of bogolons.
Since the impurity effect gives a dominant contribution at low energies, we analyzed it in detail and found that the odd-frequency pair induces the zero-energy peak in the single particle DOS in bulk.
While this analysis can capture a qualitative feature of BFS, the origin of the impurity potential on bogolons is not clear, which should be derived from the scattering potential defined in terms of original electrons.

\begin{figure*}[tb]
    \centering
    \includegraphics[width=14cm]{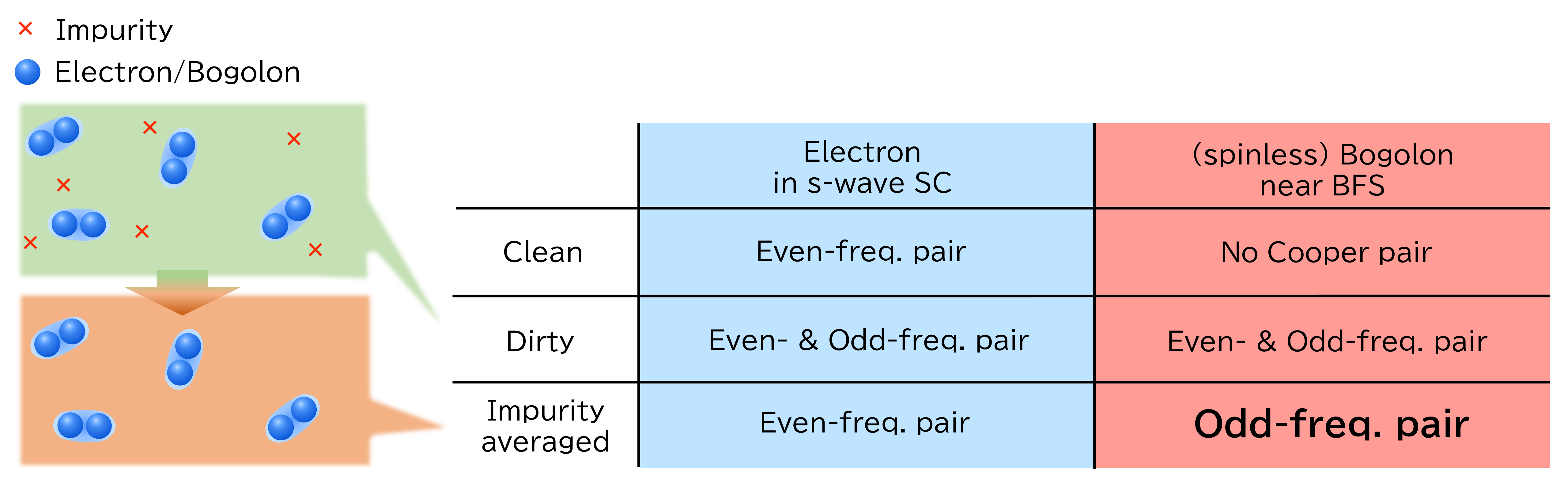}
    \caption{Schematic table for the mechanism of odd-frequency pairing of bogolons and its comparison with the $s$-wave superconductor.
    The second row (`Clean') shows the frequency dependence of the Cooper pairs in the clean limit.
    The third row (`Dirty') shows that of the dirty superconductor.
    The pairs that remain after the impurity average is indicated in the fourth row.}
    \label{fig:concept}
\end{figure*}

In this paper, we derive the effective low-energy model of bogolons by starting with the tight-binding model of $\mrm{FeSe}$.
We use the realistic tight-binding model of full $d$-orbitals at $\mrm{Fe}$ site generated from the first principle calculation.
We then put the intra-band and inter-band pair potentials following Ref.~\cite{Setty20} to create BFS.
We further consider the impurity potential defined in the original normal electrons, and clarify its effect on the bogolons near BFS.
Although these forms of the pair potential are not fully realistic, our approach can estimate the order of magnitude for physical quantities.
This work also demonstrates the validity of the phenomenological description of the low-energy physics of bogolons given in Ref.~\cite{Miki21}.

This paper is organized as follows.
In Sec.~\ref{sec:overview} we review the appearance of the odd-frequency pair of bogolons and compare it with electrons' Cooper pair.
Sections~\ref{sec:bfs} and \ref{sec:imp} are devoted to the explanation of our model of BFS based on $\mrm{FeSe}$. 
In Sec.~\ref{sec:result}, we show the numerical results for the single-particle spectra.
We summarize our result in Sec.~\ref{sec:summary}.
Below, we take the unit $\hbar = k_{\mrm{B}} = a = 1$, where $a$ is a lattice constant.
Some computational details are provided in Appendices~\ref{sec:dft}--\ref{sec:calc}.

\section{Overview: odd-frequency pairing of bogolon \label{sec:overview}}

Before we go into the details of the tight-binding model study,
we here overview the concept of our work by comparing the two cases: the electrons in the conventional ($s$-wave) superconductor and the low-energy bogolons near BFS discussed in Refs.~\cite{Miki21,Hoshino22}.
We assume that the system with BFS has the inversion symmetry and does not have time reversal symmetry.
Although the pure odd-frequency pair of bogolon is generally induced by non-ideality, i.e., disorder or interaction effects \cite{Miki21}, we below limit ourselves to the impurity effect which becomes dominant at low energies.

As schematically shown in Fig.~\ref{fig:concept}, we consider the two superconducting systems with impurity potential.
In both cases, the total Hamiltonian is given in the form $\mscr{H} = \mscr{H}_0 + \mscr{H}_{\mrm{imp}}$, where $\mscr{H}_0$ is a clean-limit part and $\mscr{H}_{\mrm{imp}}$ is an impurity potential part.
The clean-limit part is explicitly given by $\mathscr H_0 = \sum_{\bm k\sg} (\ep_{\bm k} c^\dg_{\bm k\sg} c_{\bm k\sg} + \Delta c^\dg_{\bm k\ua} c_{-\bm k,\da}^\dg + {\rm H.c.})$ for electrons with the single-particle energy $\ep_{\bm k}$, the annihilation operator $c_{\bm k\sg}$, and the $s$-wave pair potential $\Delta$ (left column).
As for the low-energy effective model of bogolons, the Hamiltonian is $\mathscr H_0 = \sum_{\bm k} \ep_{\bm k} \al^\dg_{\bm k} \al_{\bm k}$ where $\al_{\bm k}$ is the annihilation operator of bogolon (right column).
We note that the latter bogolon model describes the degrees of freedom near the BFS and does not have spin index and off-diagonal part since we assume the broken time-reversal symmetry and preserved inversion symmetry \cite{Miki21}.
Namely, we have the identity $\sum_{\bm k} \Delta_{\bm k}\al_{\bm k}^\dg \al_{-\bm k}^\dg = - \sum_{\bm k} \Delta_{-\bm k}\al_{\bm k}^\dg \al_{-\bm k}^\dg = 0$ for $\Delta_{\bm k} = \Delta_{-\bm k}$.
This point is summarized in the the second row (labeled as `Clean') of Fig.~\ref{fig:concept}.

In the presence of impurity potentials, the Green's functions $\hat G_{\bm k \bm k'}$ is written in the form of $2\times 2$ matrix by using Nambu spinor $(c_{\bm k \ua} , c_{-\bm k \da}^\dg)^\T$ for electrons with $s$-wave superconductivity and $(\al_{\bm k} , \al_{-\bm k}^\dg)^\T$ for bogolons near the BFS, respectively.
The Green's functions for each system satisfy the Dyson equation, which is written by
\begin{align}
    &\hat G_{\bm k \bm k'}(\imu\om_n) = \hat G_{\bm k}^0(\imu\om_n)\delta_{\bm k \bm k'} \nt
    &+ \hat G_{\bm k}^0(\imu\om_n) \sum_{\bm k_1} \hat u_{\bm k \bm k_1} \hat G_{\bm k_1 \bm k'}(\imu\om_n), \label{eq:dyson}
\end{align}
where $\hat u_{\bm k \bm k'}$ is a impurity scattering matrix, which is to be averaged.
$\omega_n$ is a fermionic Matsubara frequency.

One may wonder if the odd-frequency pair amplitude might not arise from the static potential $\hat u_{\bm k \bm k_1}$.
The appearance of the dynamical pair amplitude and pair potential can be understood by considering the lowest-order perturbation term.
First-order term vanishes after the random average, and hence we consider the second-order self-energy of the impurity potential:
\begin{align}
    \hat \Sigma_{\bm k \bm k'}(\imu \omega_n) = \sum_{\bm k_1} \hat u_{\bm k \bm k_1} \hat G_{\bm k_1}^0(\imu\om_n) \hat u_{\bm k_1 \bm k'}. \label{eq:bog_elec_sigma}
\end{align}
This self-energy includes both the even- and odd-frequency parts originating from the Green's function $G^0$, and breaks the inversion and translational symmetry.
This situation is summarized in the third row (labeled as `Dirty') of Fig.~\ref{fig:concept}.
After taking a impurity average, the inversion and translational symmetry are recovered (See, the fourth row of Fig.~\ref{fig:concept}.).

The above Green-function structures are same between electrons in $s$-wave superconductor and bogolons near the BFS.
Below, we clarify the difference between the two superconducting systems by focusing on the detailed structures of the Green's functions.

\subsubsection*{{\rm (a)} Electrons in $s$-wave superconductor}

First, we consider the case of the $s$-wave superconductor with inversion symmetry in clean limit.
The unperturbed Green's function is given by
\begin{align}
    \hat G_{\bm k}^0(\imu\om_n) = 
    \begin{pmatrix}
        G_{\bm k}^0(\imu\om_n) & F_{\bm k}^0(\imu\om_n) \\
        F_{\bm k}^{0\dg}(\imu\om_n) & \bar G_{\bm k}^0(\imu\om_n)
    \end{pmatrix}. \label{eq:elec_green}
\end{align}
The off-diagonal part is present already in the clean limit.
Since the impurity scattering potential is gauge invariant in terms of electrons, $\hat u_{\bm k \bm k'}$ has only the diagonal part:
\begin{align}
    \hat u_{\bm k \bm k'} = 
    \begin{pmatrix}
        u(\bm k , \bm k') & 0 \\
        0 & -u(\bm k' , \bm k)
    \end{pmatrix}. \label{eq:elec_u}
\end{align}
Inserting Eqs.~\eqref{eq:elec_green} and \eqref{eq:elec_u} into Eq.~\eqref{eq:bog_elec_sigma}, the off-diagonal self-energy, the pair potential $[\hat \Sigma_{\bm k \bm k'}(\imu\om_n)]_{12}$, can be written by
\begin{align}
    [\hat \Sigma_{\bm k \bm k'}(\imu \omega_n)]_{12} = \sum_{\bm k_1} u(\bm k , \bm k_1) F_{\bm k_1}^0(\imu\om_n) u(\bm k_1 , \bm k').
\end{align}
The frequency dependence enters through $F_{\bm k_1}^0(\imu\om_n)$ which is the even function of frequency.
Correspondingly, the pair potential also has the even frequency functional form after the impurity average [See the fourth row (labeled as `Impurity averaged') of Fig.~\ref{fig:concept}.].

\subsubsection*{{\rm (b)} Low-energy bogolons near Bogoliubov Fermi surface}

We next consider the case of bogolon.
In the clean limit, the Green's function is given by
\begin{align}
    \hat G_{\bm k}^0(\imu\om_n) = 
    \begin{pmatrix}
        G_{\bm k}^0(\imu\om_n) & 0 \\
        0 & \bar G_{\bm k}^0(\imu\om_n)
    \end{pmatrix},
    \label{eq:bog_green}
\end{align}
where the inversion symmetry prohibits the off-diagonal terms.
Since the gauge symmetry is broken, the impurity potential of bogolon generally has both diagonal ($\al^\dg \al$) and off-diagonal ($\al^\dg \al^\dg$) terms effectively, which are denoted by $u_1(\bm k, \bm k)$ and $u_2(\bm k , \bm k')$, respectively.
We write the concrete form of $\hat u_{\bm k \bm k'}$ as follows:
\begin{align}
    \hat u_{\bm k \bm k'} = 
    \begin{pmatrix}
        u_1(\bm k , \bm k') & u_2(\bm k , \bm k') \\
        u_2(\bm k' , \bm k)^\ast & -u_1(\bm k' , \bm k)
    \end{pmatrix}. \label{eq:bog_u}
\end{align}
The pair potential (=anomalous self-energy) then becomes
\begin{align}
    &[\hat \Sigma_{\bm k \bm k'}(\imu \omega_n)]_{12} = \sum_{\bm k_1} [u_1(\bm k , \bm k_1) G^0_{\bm k_1}(\imu\om_n) u_2(\bm k_1 , \bm k') 
    \nt
    &\hspace{25mm} + u_2(\bm k , \bm k_1) G^0_{\bm k_1}(\imu\om_n) u_1(\bm k_1 , \bm k')].
\end{align}
We note that the diagonal Green's function $G_{\bm k_1}^0(\imu\om_n)$ is composed of mixed even and odd functions of frequency. 
Since the pair potential obeys the Fermi-Dirac statistics, $[\hat \Sigma_{\bm k \bm k'}]_{12}$ becomes the odd function of frequency in inversion-symmetric system after the impurity average (the fourth row of Fig.~\ref{fig:concept}).
The pure odd-frequency pair of bogolon is thus realized.
We note that the interaction effect also induces the self-energies and the odd-frequency pairing amplitude, although the impurity effect is dominant at low-frequency and low-temperature limit \cite{Miki21}.

In the following sections, in order to clarify the microscopic origin of $u_1$ and $u_2$ and their physical consequences, we investigate the impurity effect on BFS based on the realistic tight-binding model, where the impurity potential is defined in the real-space representation in terms of the original electrons.

\section{Model Hamiltonian and Bogoliubov Fermi Surface \label{sec:bfs}}

\begin{figure}[tb]
    \centering
    \includegraphics[width=8.5cm]{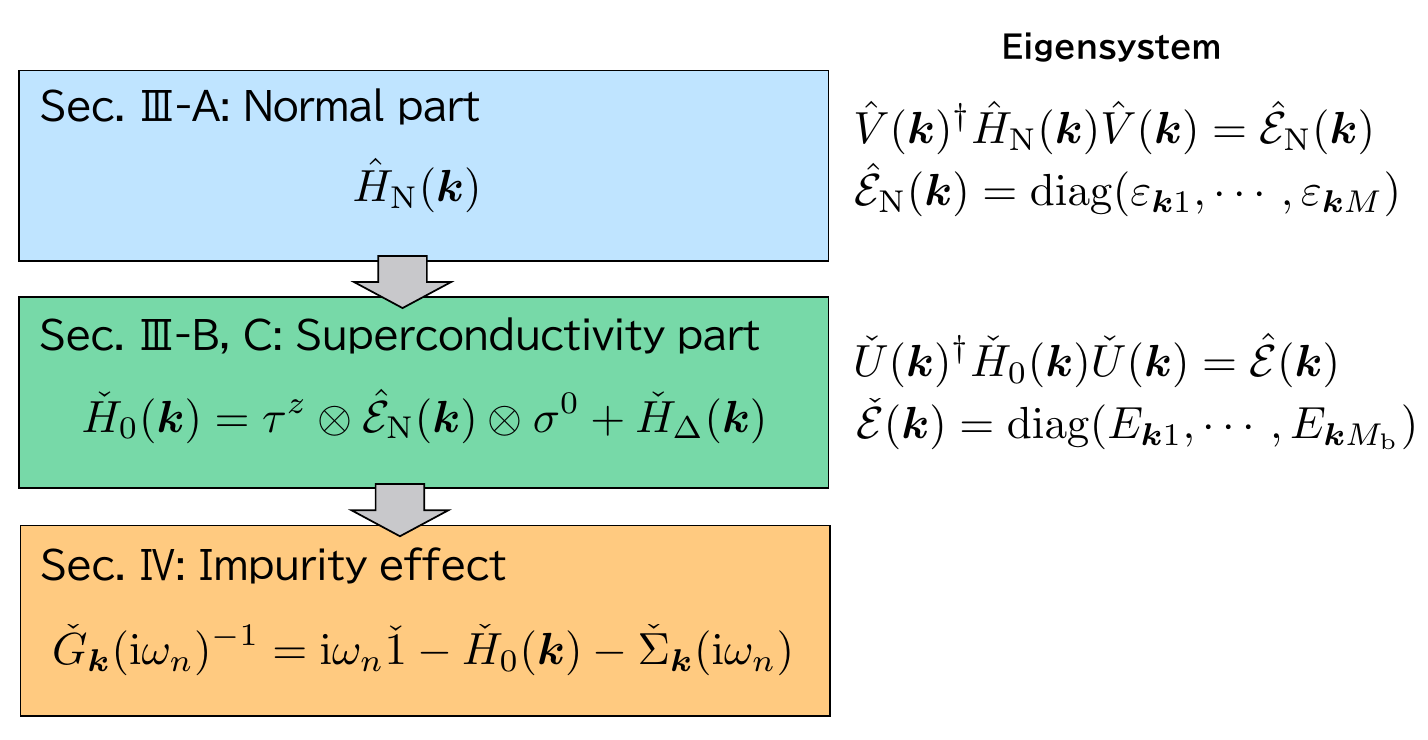}
    \caption{
    Flow of the calculation.
    The normal part is discussed in Sec.~\ref{sec:bfs}-A, while Sec.~\ref{sec:bfs}-B and -C deal with the superconductivity part. 
    We list the notations of the Hamiltonians, the eigenenergies and vectors for each part as shown in the figure.
    We discuss the impurity effect in Sec.~\ref{sec:imp}.}
    \label{fig:flow}
\end{figure}

In this section, we introduce a model for BFS.
The total Hamiltonian is composed of the three parts: $\mscr{H} = \mscr{H}_{\mrm{N}} + \mscr{H}_{\Delta} + \mscr{H}_{\mrm{imp}}$, where $\mscr{H}_{\mrm{N}}$ is a normal state Hamiltonian for the clean limit, $\mscr{H}_{\Delta}$ is a pair potential part, and $\mscr{H}_{\mrm{imp}}$ is a impurity potential part.
These are respectively discussed in Secs.~\ref{sec:bfs}-A, \ref{sec:bfs}-B, and \ref{sec:imp}.
Since the procedure is complicated, we summarize the calculation flow and notations in Fig.~\ref{fig:flow}.

\subsection{Normal state for clean limit}

Below, we construct the Hamiltonian with the real material $\mrm{Fe(Se,S)}$ in mind.
Since the effect of the S-doping is expected to give a chemical pressure, we assume that it does not change the band structure significantly.
Accordingly, we use the tight binding parameters based on the first principle calculation of $\mrm{FeSe}$.
However, as is well known, the number and size of the experimentally observed Fermi surfaces deviate significantly from the first-principles calculations \cite{Yamakawa16}, so we have adjusted their band structure here (See Appendix A for details.).
Although $\mrm{FeSe}$ has a nematic transition from the tetragonal ($P4/nmm$) phase to the orthorhombic ($Cmma$) phase, we use the hopping parameters for the tetragonal case.
This is because the finite zero-energy DOS inside the superconducting phase is observed in the tetragonal phase of $\mrm{Fe(Se,S)}$ experimentally \cite{Hanaguri18, Sato18, Mizukami21}. 
Below we do not consider the spin-orbit coupling in the normal state for simplicity.

From the band-structure calculation, we can obtain the tight-binding Hamiltonian written by the orthogonal basis of the Wannier function $w_\gm(\bm r - \bm R_{\bm n} - \bm d_a)$, where
$\bm R_{\bm n}$ denotes the center of each unit cell, and $\bm d_a\, (a = \mrm{Fe}\, 1, \mrm{Fe}\, 2)$ specifies the position of $\mrm{Fe}$ inside the unit cell measured from $\bm R_{\bm n}$.
The atomic orbitals are described by $\gm = z^2, xz, yz, x^2-y^2, xy$.
See Appendix~\ref{sec:dft} for more details about the derivation of tight-binding model parameters.
The normal state Hamiltonian is written by the creation and annihilation operators as
\begin{align}
  &\mscr{H}_{\mrm{N}} = \sum_{\bm n \bm m} \sum_{a a' \gm \gm' \sg \sg'} H_{\mrm N \gm \gm'}(\bm R_{\bm n} + \bm d_{a'} - \bm d_a) \nt 
  &\times c_{\gm \sg}^\dg(\bm R_{\bm m} + \bm d_{a} ) c_{\gm' \sg}(\bm R_{\bm m} + \bm R_{\bm n} + \bm d_{a'}) \nt
  &- \mu \sum_{\bm n a \gm \sg} c_{\gm \sg}^\dg(\bm R_{\bm n} + \bm d_{a} ) c_{\gm \sg}(\bm R_{\bm n} + \bm d_{a}). \label{eq:ham_n}
\end{align}
Since the original parameters have the numerical errors, we use the parameters averaged by the symmetry operation as explained in Appendix~\ref{sec:symmetry}.

\begin{figure}[tb]
    \centering
    \includegraphics[width=8.5cm]{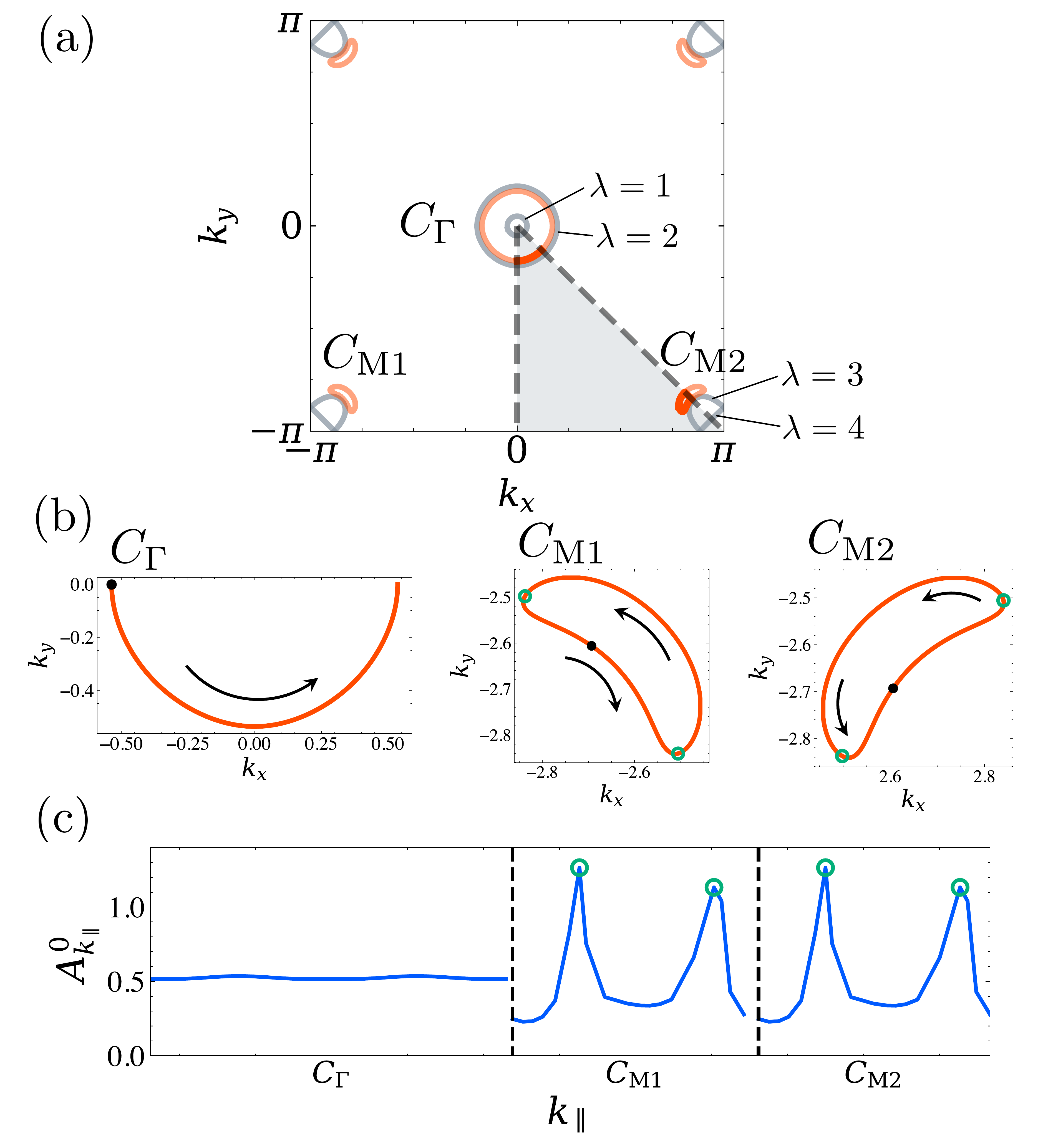}
    \caption{(a) Normal Fermi surfaces (gray) and BFSs (red). The magnified view of the BFSs are shown in (b).
    (c) Zero-energy spectral function for clean limit. The horizontal axis labeled as $k_\parallel$ is path shown by the arrows in (b), in which the beginning of the path is shown by black point. 
    The characteristic $\bm k$ points are indicated by green circles in (b), and the corresponding $\bm k$ points are shown by the same symbols in (b).
    The number of $k_\parallel$ mesh in (c) is $88$ on the BFSs. The energy unit is taken as $\mrm{eV}$.}
    \label{fig:fermi}
\end{figure}

Performing the Fourier transformations with respect to the lattice vector $\bm R_{\bm n}$, we obtain the Fourier component of the Hamiltonian expressed by $H_{\mrm N a\gm , a' \gm'}(\bm k)$.
Then we diagonalize the matrix $H_{\mrm N a\gm , a' \gm'}(\bm k)$ at each $\bm k$ point, and calculate the electron band energies numerically.
We express this as
\begin{align}
    \hat V(\bm k)^\dg \hat H_{\mrm N}(\bm k) \hat V(\bm k) &= \hat{\mcal{E}}_{\mrm{N}}(\bm k) \nt 
    &= \mrm{diag}\, (\ep_{\bm k 1}, \cdots , \ep_{\bm k M}), \label{eq:e_normal}
\end{align}
where the hat ($\hat \ $) symbol represents a $M\times M$ matrix where $M=\sum_{a\gm}1 = 10$.
Then, we obtain the diagonalized 
Hamiltonian as
\begin{align}
    \mscr{H}_{\mrm{N}} = \sum_{\bm k} \sum_{\lam \sg} (\ep_{\bm k\lam} - \mu) c_{\bm k\lam \sg}^\dg c_{\bm k\lam \sg}, \label{eq:ham_nband}
\end{align}
where $\lam$ is a band index.
The Fermi surfaces in the normal state is shown in Fig.~\ref{fig:fermi} (a) with gray solid lines.
The two hole pockets appears around the $\Gamma$ point, and the electron pocket around the $\mrm M$ point.

In order to introduce the pair potential in the band basis, we need to identify which bands are connected to each other at different $\bm k$ points, especially when two bands are crossed.
For this identification, we consider the eigenvector $\bm v(\bm k)_{\lam}$ written as
\begin{align}
    \bm v(\bm k)_{\lam} = ([\hat V(\bm k)]_{a_1\gm_1 , \lam} , \cdots , [\hat V(\bm k)]_{a_M \gm_M , \lam})^\T. \label{eq:eigenvec}
\end{align}
We determine which band ($\lambda'$) at $\bm k + \Delta\bm k$ connect to the band $\lam$ at $\bm k$ by using the inner product of the eigenvectors.
Namely, for a given $\bm v(\bm k)_{\lambda}$, we choose the index $\lambda'$ which maximizes the magnitude of $|\bm v(\bm k + \Delta\bm k)_{\lam'} \cdot \bm v(\bm k)_{\lam}|$.
Thus we make the band structure smooth for each band index.

\subsection{Pair potential part}

Here, we consider the pair potential part phenomenologically, following the procedure given in Ref.~\cite{Setty20}.
It is important to consider the inter-band pairing with time reversal symmetry breaking for BFS.
Below, for convenience of explanation, we assume that the bands with the indices $\lam=1 , 2 , 3 , 4$ constitute the normal Fermi surfaces, which are shown in Fig.~\ref{fig:fermi} (a) with gray lines.
$\lam = 1 , 2$ corresponds to the two Fermi surfaces around the $\Gamma$ point, where the small one is labeled as $\lam=1$ and the large one as $\lam=2$.
Similarly, $\lam = 3 , 4$ makes the Fermi surfaces around the $\mrm M$ point (the curved one at the corner is labeled as $\lam=3$, and the straight one as $\lam=4$).

The pair potential term is written in the form
\begin{align}
  &\mscr{H}_{\Delta}
  = \sum_{\bm k} \sum_{\sg\sg'} [(\delta \sg^x + \imu \Delta_0 \sg^y)\imu \sg^y]_{\sg\sg'} \nt 
  &\times \Bigg( \sideset{}{^{'}}\sum_{\lam < \lam'} c_{\bm k\lam \sg}^\dg c_{-\bm k\lam' \sg'}^\dg
  -\sideset{}{^{'}}\sum_{\lam > \lam'} c_{\bm k \lam' \sg}^\dg c_{-\bm k \lam \sg'}^\dg \Bigg) + \Hc \nt
  &+ \sum_{\bm k} \sum_{\lam \in \mrm{FS}} \Delta_\lam (\bm k) c_{\bm k\lam \ua}^\dg c_{-\bm k\lam \da}^\dg + \Hc, \label{eq:ham_delta}
\end{align}
where 
the summation of $\lam \in \mrm{FS}$ is taken over $\mrm{FS} = \{1,2,3,4\}$, which reflects the fact that the electrons constituting the Fermi surfaces participate in the Cooper-pair condensation.
$\sideset{}{^{'}}{\textstyle\sum_{\lam \gtrless \lam'}}$
indicates that the summation is taken if at least one of the bands $\lam$ and $\lam'$ has a Fermi surface.
$\Delta_0, \delta$ are inter-band pair potentials, and $\Delta_{\lam}(\bm k)$ is a intra-band pair potential.
$\sg^x$ and $\sg^y$ are $x$ and $y$ components of the Pauli matrices, respectively.
Note that inter-band pairing term breaks the time reversal symmetry, because this term has a form of $\sim \sg^x + i \sg^y$.
Since these microscopic parameters are not explicitly known,
we take the extended $s$-wave intra-band pair potential $\Delta_\lam(\bm k) = \Delta_{0\lam} + \Delta_{1\lam} \cos(k_x/2)\cos(k_y/2)\, (\lam \in \mathrm{FS})$ for a concrete calculation and for estimation of the order of magnitude of physical quantities.
Each parameter is chosen as follows: $\delta = \Delta_0 = 0.03, \Delta_{0,\lam=1} = 0.03, \Delta_{0,\lam=2} = 0.07, \Delta_{0,\lam=3} = 0.05, \Delta_{0,\lam=4} = 0.05, \Delta_{1,\lam=1} = 0.01, \Delta_{1,\lam=2} = 0.01, \Delta_{1,\lam=3} = -0.02, \Delta_{1,\lam=4} = -0.02$ (The unit is given in $\mrm{eV}$.).
The resulting BFSs [shown in Fig.~\ref{fig:fermi} (a)] resemble the one shown in Ref.~\cite{Setty20}.

We define the Nambu spinor by $\vec \Psi_{\bm k} = (c_{\bm k 1 \ua}, c_{\bm k 1 \da}, c_{\bm k 2 \ua}, \cdots c_{\bm k M \da} , c_{-\bm k 1 \ua}^\dg , \cdots , c_{-\bm k M \da}^\dg)^\T$, with $M_{\mrm{b}} = 2 \times 2 \times M$ components (each of which corresponds to Nambu, spin, and band spaces).
Then, the BdG Hamiltonian is written as
\begin{align}
    \mscr{H}_0 &= \mscr{H}_{\mrm{N}} + \mscr{H}_{\Delta} \nt
    &= \sum_{\bm k \in \mrm{HBZ}} \vec \Psi_{\bm k}^\dg \check H_0(\bm k) \vec \Psi_{\bm k}, \label{eq:ham_total}
\end{align}
where the summation of $\bm k$ is taken over the half Brillouin zone (HBZ, $k_y < 0$).
We introduced the matrix representation of the Hamiltonian at $\bm k$ as $\check H_0(\bm k) = \tau^0 \otimes \hat H_{\mrm N}(\bm k) \otimes \sg^0 + \check H_{\Delta}(\bm k)$, where $\check H_{\Delta}(\bm k)$ is the matrix form of the pair potential part defined in Eq.~\eqref{eq:ham_delta} in Nambu basis, and $\tau^0, \sg^0$ are two dimensional identity matrices in Nambu space and spin space, respectively.
We also define the unitary transformation to the bogolon basis as
\begin{align}
    [\vec \Psi_{\bm k}]_j = \sum_{b} [\check U(\bm k)]_{j b} \vec \al_{\bm k b}. \label{eq:u}
\end{align}
where $\vec \al_{\bm k} = (\al_{\bm k 1}, \cdots, \al_{\bm k 2M} , \al_{-\bm k 1}^\dg , \cdots , \al_{-\bm k 2M}^\dg)^\T$.
$j = 1, \cdots, M_{\mrm{b}}$ is a index of the spinor in electron basis, while $b = 1, \cdots , M_{\mrm{b}}$ is a index of spinor in bogolon basis.
Then, Eq.~\eqref{eq:ham_total} can be rewritten as
\begin{align}
    \mscr{H}_0 &= \sum_{\bm k \in \mrm{HBZ}} \vec \al_{\bm k}^\dg \check{\mcal{E}}(\bm k) \vec \al_{\bm k},
\end{align}
where 
\begin{align}
    \check U(\bm k)^\dg \check H_0(\bm k) \check U(\bm k) 
    &= \check{\mcal{E}}(\bm k) \nt
    &= \mrm{diag}\, (E_{\bm k 1} , \cdots, E_{\bm k M_{\mrm{b}}}). \label{eq:e_super}
\end{align}
Figure~\ref{fig:fermi} (a) shows the BFSs (red lines) together with the normal Fermi surfaces (gray solid lines).
The magnified view of the BFSs are shown in (b), and the $\bm k$-dependent spectra is also shown in (c). 
Since the BFSs are topologically protected, we have identified the Bogoliubov Fermi wave-vector by the sign change of the Pfaffian $\mrm{Pf}(\bar{H}_0(\bm k))$ for the antisymmetrized Hamiltonian \cite{Agterberg17, Brydon18, Setty20, Pfaff}
\begin{align}
    \bar{H}_0(\bm k) = \check W \check H_0(\bm k) \check W^\dg,
\end{align}
where
\begin{align}
    \check W = 
    \frac{1}{\sqrt{2}}
    \begin{pmatrix}
        1 & 1 \\
        \imu & -\imu
    \end{pmatrix} \otimes I \otimes \sg^0.
\end{align}
$I$ is the identity matrices with the dimension $M$ (band space).
We note that the eigenvalues $E_{\bm k b}$ of the Kramers pair states are not degenerate because of the time-reversal symmetry broken pair potential in Eq.~\eqref{eq:ham_delta}.
Therefore, there are two bands of bogolon $b = 1,2$ crossing at the Fermi level (One is particle band and the other is anti-particle band), which constitute the BFSs.

\subsection{Phase of wave-function at each $\bm k$\label{sec:phase}}

Since the pair potential is defined in the band basis ($\lambda$), the phase of the eigenfunctions must be determined with careful thoughts.
Even after ordering $\lam$ in each $\bm k$ point using  the eigenvectors Eq.~\eqref{eq:eigenvec}, there are still U(1) gauge degrees of freedom.
Namely, the transformation $\bm v(\bm k)_{\lam} \to \epn^{\imu \theta_\lam(\bm k)} \bm v(\bm k)_{\lam}$ does not alter the energy-eigenvalues Eq.~\eqref{eq:e_normal}.
This degrees of freedom are usually not reflected in physical quantity.
However the spectral functions in the superconducting states is in general not invariant under this transformation, because our pair potential part $\mscr{H}_\Delta$ is given in band basis, while the normal part $\mscr{H}_{\mrm{N}}$ is given in the basis of the Wannier function.

As will be discussed in Sec.~\ref{sec:result}, we will concentrate on the low-energy contribution near the BFSs, and hence it is necessary to construct a smooth function for the paths along BFSs $C_\Gamma, C_{\mrm M1}, C_{\mrm M2}$ [shown in Fig.~\ref{fig:fermi} (a) and (b)].
We also choose the phase so as to preserve the inversion symmetry with which the theoretical results are consistent with the experiment \cite{Nagashima22}.
More specifically, we consider a (non-twisted) parallel-transport gauge for $C_{\mrm M1}, C_{\mrm M2}$ and a twisted parallel-transport gauge with inversion symmetry for $C_{\Gamma}$ \cite{Vanderbilt_book}.
Let us write the eigenvector with the parallel-transport gauge as $\bar{\bm v}(\bm k)$ and the numerically obtained one as $\bm v_{\mrm{num}}(\bm k)$.
We fix the phase at the beginning point $\bm k_0$ for each path as $\bar{\bm v}(\bm k_0)_{\lam} = \bm v(\bm k_0)_{\lam}$ [indicated by the black dot in Fig.~\ref{fig:fermi} (b)].
The phase of the next $\bm k$-point is chosen as it is parallel to previous point.
For this purpose, we first introduce the relative phase between $\bm k$ and $\bm k + \Delta \bm k$ as
\begin{align}
    \varphi_\lam(\bm k) = -\imag \ln \bar{\bm v}(\bm k)_{\lam} \cdot \bm v_{\mrm{num}}(\bm k + \Delta \bm k)_{\lam}. \label{eq:varphi}
\end{align}
This quantity corresponds to the Berry connection in a continuous limit.
Then we define the eigenvector with parallel-transport gauge at $\bm k + \Delta \bm k$ as
\begin{align}
    \bar{\bm v}(\bm k + \Delta \bm k)_{\lam} = \epn^{\imu\varphi_\lam(\bm k)}\bm v_{\mrm{num}}(\bm k + \Delta \bm k)_{\lam}.
\end{align}
In this gauge, $\bar{\bm v}(\bm k)_{\lam}$ and $\bar{\bm v}(\bm k + \Delta \bm k)_{\lam}$ become parallel:
\begin{align}
    -\imag \ln \bar{\bm v}(\bm k)_{\lam} \cdot \bar{\bm v}(\bm k + \Delta \bm k)_{\lam} = 0.
\end{align}
The Berry phase $\phi_C$ can be calculated numerically by the summation of the left hand side of the above equation taken over the the closed paths $C = C_{\mrm M1}, C_{\mrm M2}$.
We obtain $\phi_{C} = 0$ because of the inversion symmetry and time reversal symmetry of the normal state.
For their counter parts ($k_y > 0$), we can obtain the smooth and inversion symmetric eigenvectors by using the symmetry operation defined in Eq.~\eqref{eq:eigenvec_sym}.

For unclosed paths $C_{\Gamma}$ [See, Fig.~\ref{fig:fermi} (b)], we choose the twisted parallel transport gauge $\tilde{\bm v}(\bm k)$ form $\bar{\bm v}(\bm k)$.
In numerical calculation, we calculate the $\bar{\bm v}(\bm k_0)_\lam, \cdots, \bar{\bm v}(\bm k_{N_\Gamma - 1})_\lam$ on the path $C_\Gamma$, and determine $\bar{\bm v}(-\bm k_0)_\lam$ by using inversion symmetric operation for $\bar{\bm v}(\bm k_0)_\lam$ [See Eq.~\eqref{eq:eigenvec_sym} in Sec.~\ref{sec:summary}.].
Then, we calculate the relative phase between $\bar{\bm v}(\bm k_{N_\Gamma - 1})_\lam$ and $\bar{\bm v}(-\bm k_0)_\lam$, which is written by
\begin{align}
    \varphi_{\lam}' = -\imag \ln \bar{\bm v}(\bm k_{N_\Gamma - 1})_\lam \cdot \bar{\bm v}(-\bm k_0)_\lam.
\end{align}
Finally, we define the twisted parallel transport for $n$-th point ($n = 0, \cdots , N_\Gamma - 1$) of the path by twisting the phase at each point
\begin{align}
    \tilde{\bm v}(\bm k_n)_\lam = \epn^{\imu \varphi_{\lam}' n /N_\Gamma} \bar{\bm v}(\bm k_n)_\lam.
\end{align}
This choice of gauge results in smooth and inversion-symmetry-preserved eigenvector.

\section{Impurity effects and Green's functions \label{sec:imp}}

In this section, we proceed to a concrete analysis of the impurity effects.
We use the Green's function method, which is appropriate for the analysis of the impurity effect on the superconducting system \cite{Abrikosov59, Abrikosov61}.
We calculate the spectral function which is an experimentally observable quantity.

\subsection{Definition of Green's functions and self-energies}

First, we define the Green's function as
\begin{align}
    \check G_{\bm k}(\tau) = -\la \mcal{T} \vec \Psi_{\bm k}(\tau) \vec \Psi_{\bm k}^\dg \ra
    = \begin{pmatrix}
    G_{\bm k}(\tau) & F_{\bm k}(\tau) \\
    F^\dg_{\bm k}(\tau) & \bar G_{\bm k}(\tau)
    \end{pmatrix}
    , \label{eq:green}
\end{align}
where $\mcal{T}$ represents imaginary time ordering, $\la \cdots \ra$ is a statistical average, and $A(\tau) = \epn^{\tau \mscr{H}} A \epn^{-\tau \mscr{H}}$ is the Heisenberg representation with imaginary time. 
$G_{\bm k},F_{\bm k}$ correspond to the normal and anomalous Green's functions, respectively, and their conjugate quantities are $\bar G_{\bm k},F_{\bm k}^\dg$.
The Fourier transformation from imaginary time to Matsubara frequency is defined by
\begin{align}
    \check G_{\bm k}(\imu\om_n) = \int_0^{1/T} \diff\tau\, \check G_{\bm k}(\tau) \epn^{\imu\om_n \tau},
\end{align}
where $T$ is a temperature.
Using the clean limit Hamiltonian $\check H_0(\bm k)$ in Eq.~\eqref{eq:ham_total}, the self-energy is also introduced by  
\begin{align}
    \check G_{\bm k}(\imu\om_n)^{-1} = \imu \om_n \check 1 - \check H_0(\bm k) - \check \Sigma_{\bm k}(\imu\om_n). \label{eq:def_selfenergy}
\end{align}
Each component of the self-energy is expressed as 
$\displaystyle \check \Sigma_{\bm k} = 
\begin{pmatrix}
\Sigma_{\bm k} & S_{\bm k} \\
S_{\bm k}^\dg & \bar{\Sigma}_{\bm k}
\end{pmatrix}
$.

\subsection{Impurity potential}

We now explain the impurity potential part.
It is convenient to deal with real space representation for the disorder potential.
We start from the expression written by the field operator $\psi , \psi^\dg$ as follows:
\begin{align}
  \mscr{H}_{\mrm{imp}} &= \sum_i \sum_{\sg \sg'} \int \diff \bm r\, \psi_\sg^\dg(\bm r) U_{\mrm{imp}}^{\sg \sg'}(\bm r - \bm r_i) \psi_{\sg'}(\bm r), \label{eq:ham_imp}
\end{align}
where $U_{\mrm{imp}}^{\sg \sg'}(\bm r - \bm r_i)$ is an impurity potential with spin $\sg, \sg'$, and $\bm r_i$ is a scattering center.
We can move to the Wannier function basis from Eq.~\eqref{eq:ham_imp} by expanding the field operator, which is written by
\begin{align}
    \psi_\sg(\bm r) = \sum_{\bm n} \sum_{a \gm} w_\gm(\bm r - \bm R_{\bm n} - \bm d_a) c_{\gm \sg}(\bm R_{\bm n} + \bm d_a). \label{eq:psi}
\end{align}
Inserting Eq.~\eqref{eq:psi} into Eq.~\eqref{eq:ham_imp}, we obtain the impurity potential part of the Hamiltonian
\begin{align}
  &\mscr{H}_{\mrm{imp}} =  \int \frac{\diff \bm q}{(2\pi)^3} \sum_{\bm k , \bm k'} \sum_{\lam , \lam' , \sg , \sg'} \sum_{a a' \gm \gm' \sg \sg'} \rho_{\bm q} U_{\mrm{imp}}^{\sg \sg'}(\bm q) \nt 
  &\times f_{a\gm, a'\gm'}^{\bm k \bm k'}(\bm q) [\hat V(\bm k)^\dg]_{\lam, a \gm} [\hat V(\bm k')]_{a' \gm' , \lam'} c_{\bm k \lam \sg}^\dg c_{\bm k' \lam' \sg'}, \label{eq:ham_imp_k}
\end{align}
where the form factor $f_{a \gm , a' \gm'}^{\bm k \bm k'}(\bm q)$ can be written as
\begin{align}
    &f_{a \gm , a' \gm'}^{\bm k \bm k'}(\bm q) = 
    \frac{1}{N}\sum_{\bm n , \bm m} \int \diff\bm r w_{\gm}^\ast(\bm r - \bm R_{\bm n} - \bm d_a) \nt  
    &\times w_{\gm'}(\bm r - \bm R_{\bm m} - \bm d_{a'}) \epn^{\imu \bm q \cdot \bm r -\imu \bm k\cdot \bm R_{\bm n} + \imu \bm k' \cdot \bm R_{\bm m}},
\end{align}
and $\rho_{\bm q} = \sum_i \epn^{-\imu\bm q \cdot \bm r_i}$ is a structure factor for the impurity configuration $\bm r_i$.
Note that $\bm q$ is defined in an infinite range, while $\bm k$ is defined in the Brillouin zone.

Now we apply the random average for impurity configuration
\begin{align}
    \overline{\rho_{\bm q} \rho_{\bm q'}} 
    = V n_{\mrm{imp}} \delta_{\bm q , -\bm q'},
\end{align}
where $n_{\mrm{imp}} = V^{-1}\sum_i 1$ ($V$ is a system volume).
We consider the second-order self-energy with respect to the impurity potential as follows
\begin{align}
    &\Sigma_{\bm k \lam \sg , \lam' \sg'}(\imu \om_n) = \frac{1}{N}
    \sum_{\bm k_1 \in {\rm HBZ}} \sum_{\lam_1 \lam_2 \sg_1 \sg_2} \nt
    &\times \Bigg[ \mathscr{U}_{\lam \sg , \lam_1 \sg_1 , \lam_2 \sg_2 , \lam' \sg'}(\bm k , \bm k_1 , \bm k_1 , \bm k) G_{\bm k_1 \lam_1 \sg_1 , \lam_2 \sg_2}(\imu\om_n) \nt 
    &- \mathscr{U}_{\lam \sg , \lam_1 \sg_1 , \lam_2 \sg_2 , \lam' \sg'}(\bm k , -\bm k_1 , -\bm k_1 , \bm k) \bar G_{\bm k_1 \lam_2 \sg_2 , \lam_1 \sg_1}(-\imu\om_n) \Bigg], \label{eq:sg} \\
    &S_{\bm k \lam \sg , \lam' \sg'}(\imu \om_n) =
    - \frac{1}{N} \sum_{\bm k_1 \in {\rm HBZ}} \sum_{\lam_1 \lam_2 \sg_1 \sg_2} \nt 
    &\times \Bigg[\mathscr{U}_{\lam_1 \sg_1 , \lam \sg , \lam' \sg' , \lam_2 \sg_2}(\bm k , \bm k_1 , -\bm k , -\bm k_1) F_{\bm k_1 \lam_1 \sg_1 , \lam_2 \sg_2}(\imu\om_n) \nt 
    &-\mathscr{U}_{\lam_1 \sg_1 , \lam \sg , \lam' \sg' , \lam_2 \sg_2}(\bm k , -\bm k_1 , -\bm k , \bm k_1) F_{\bm k_1 \lam_2 \sg_2 , \lam_1 \sg_1}(-\imu\om_n) \Bigg]
    , \label{eq:s}
\end{align}
where 
\begin{align}
  &\mathscr{U}_{\lam_1 \sg_1 , \lam_2 \sg_2 , \lam_3 \sg_3 , \lam_4 \sg_4}(\bm k_1, \bm k_2, \bm k_3, \bm k_4) \nt
  &= n_{\mrm{imp}} N \int \frac{\diff\bm q}{(2\pi)^3} \sum_{a_1 \cdots a_4} \sum_{\gm_1 \cdots \gm_4} U_{\mrm{imp}}^{\sg_1 \sg_2}(\bm q) U_{\mrm{imp}}^{\sg_3 \sg_4}(-\bm q) \nt
  &\times  [\hat V(\bm k_1)^\dg]_{\lam_1 , a_1 \gm_1} [\hat V(\bm k_2)]_{a_2 \gm_2, \lam_2} [\hat V(\bm k_3)^\dg]_{\lam_3 , a_3 \gm_3} [\hat V(\bm k_4)]_{a_4 \gm_4, \lam_4} \nt
  &\times 
  f_{a_1\gm_1, a_2\gm_2}^{\bm k_1 \bm k_2}(\bm q) f_{a_3\gm_3, a_4\gm_4}^{\bm k_3 \bm k_4}(-\bm q) \delta_{\bm k_1 + \bm k_3 , \bm k_2 + \bm k_4}. \label{eq:scr_u}
\end{align}
We also obtain $\bar \Sigma_{\bm k \lam \sg , \lam' \sg'}(\imu \om_n)$ and $S^\dg_{\bm k \lam \sg , \lam' \sg'}(\imu \om_n)$ in a similar manner to the above expressions.
We note that $\delta_{\bm k_1 + \bm k_3 , \bm k_2 + \bm k_4}$ in \eqref{eq:scr_u} originates from the recovered-translational symmetry after the random average.
If we regard $G, F, \bar G, F^\dg$ as unperturbed Green's functions, we obtain the self-energy of Born approximation.
On the other hand, if we regard them as dressed Green's functions, we obtain the self-energy of self-consistent Born approximation.

\subsection{Evaluation of self-energies}

In order to evaluate the self-energies, we need to define the specific form of the impurity potential $U_{\mrm{imp}}(\bm q)$.
First, we consider the case of non-magnetic impurity (later we will also discuss the case of magnetic impurity).
The impurity potential is given by
\begin{align}
    U_{\mrm{imp}}^{\sg \sg'}(\bm q) = U_{\mrm{imp}} \delta_{\sg \sg'},
\end{align}
which is $\bm q$-independent and is frequently used for an electron gas model.
$U_{\mrm{imp}}$ is a magnitude of the potential.
Although this is not a realistic impurity potential, 
we can further analyze the model in a simple form and make a semi-quantitative estimate of the effect of impurities.

The concrete form of the form factor in Eq.~\eqref{eq:scr_u} is written using the Wannier functions as
\begin{align}
  &\int \frac{\diff\bm q}{(2\pi)^3}\, f_{a_1\gm_1, a_2\gm_2}^{\bm k_1 \bm k_2}(\bm q) f_{a_3\gm_3, a_4\gm_4}^{\bm k_3 \bm k_4}(-\bm q) \delta_{\bm k_1 + \bm k_3 , \bm k_2 + \bm k_4} \nt
  &= \frac{1}{N^2} \sum_{\bm n_1 , \cdots , \bm n_4} \int \diff\bm r\, w_{\gm_1}^\ast(\bm r - \bm R_{\bm n_1} - \bm d_{a_1}) \nt 
  &\times w_{\gm_2}(\bm r - \bm R_{\bm n_2} - \bm d_{a_2}) w_{\gm_3}^\ast(\bm r - \bm R_{\bm n_3} - \bm d_{a_3}) \nt
  &\times w_{\gm_4}(\bm r - \bm R_{\bm n_4} - \bm d_{a_4}) \nt 
  &\times \epn^{-\imu\bm k_1 \cdot \bm R_{\bm n_1} + \imu\bm k_2 \cdot \bm R_{\bm n_2} - \imu\bm k_3 \cdot \bm R_{\bm n_3} + \imu\bm k_4 \cdot \bm R_{\bm n_4}} \delta_{\bm k_1 + \bm k_3 , \bm k_2 + \bm k_4},
\end{align}
where we have performed $\bm q$ integration.
To proceed the calculation further, we use the two approximations.
First, we observe that the above quantity is expected to become largest when the locality condition $\bm R_{\bm n_1} = \bm R_{\bm n_2} = \bm R_{\bm n_3} = \bm R_{\bm n_4}, a_1 = a_2 = a_3 = a_4$ is satisfied. 
Hence, we assume that the integration of $\bm r$ takes finite value only if it satisfies this condition.
Second, we replace the Wannier function $w_{\gm}(\bm r - \bm R_{\bm n} - \bm d_a)$ with atomic orbital function $\phi_{\gm}(\bm r - \bm R_{\bm n} - \bm d_a)$ for simplicity.
We write the atomic orbital function as $\phi_{\gm}(\bm r) = R(r) \Theta_{\gm}(\theta , \varphi)$, where $R(r)$ is a radial wave function and $\Theta(\theta, \varphi)$ is a cubic harmonics for the $d$-orbital, whose specific form is shown in Appendix~\ref{sec:cubic}.
Then we can evaluate  the integral with respect to $\theta, \varphi$ by using Eq.~\eqref{eq:cubit_int}, and obtain
\begin{align}
    &\mathscr{U}_{\lam_1 \sg_1 , \lam_2 \sg_2 , \lam_3 \sg_3 , \lam_4 \sg_4}(\bm k_1, \bm k_2, \bm k_3, \bm k_4) \nt
    &= \Gamma \mathscr{F}_{\lam_1 \sg_1, \lam_2 \sg_2, \lam_3 \sg_3, \lam_4 \sg_4}(\bm k_1 , \bm k_2 , \bm k_3 , \bm k_4), \label{eq:uf}
\end{align}
where
\begin{align}
  &\Gamma = \frac{5 n_{\mrm{imp}} U_{\mrm{imp}}^2}{28\pi} \int \diff r\, r^2 |R(r)|^4 
\end{align}
and
\begin{align}
  &\mscr{F}_{\lam_1 \sg_1, \lam_2 \sg_2, \lam_3 \sg_3, \lam_4 \sg_4}(\bm k_1 , \bm k_2 , \bm k_3 , \bm k_4) \nt
  &= \sum_{a_1 \cdots a_4} \sum_{\gm_1 \cdots \gm_4} [\hat V(\bm k_1)^\dg]_{\lam_1 , a_1 \gm_1} [\hat V(\bm k_2)]_{a_2 \gm_2 ,\lam_2} \nt
  &\times [\hat V(\bm k_3)^\dg]_{\lam_3, a_3 \gm_3} [\hat V(\bm k_4)]_{a_4 \gm_4, \lam_4} \nt
  &\times (\delta_{\gm_1 , \gm_2} \delta_{\gm_3 , \gm_4} + \delta_{\gm_1 , \gm_3} \delta_{\gm_2 , \gm_4} + \delta_{\gm_1 , \gm_4} \delta_{\gm_2 , \gm_3}) \nt
  &\times \delta_{a_1 , a_2} \delta_{a_2 , a_3} \delta_{a_3 , a_4} \delta_{\sg_1 \sg_2} \delta_{\sg_3 \sg_4}. \label{eq:scrf}
\end{align}
Once this form factor is obtained, we can immediately evaluate the self-energies by using Eqs.~\eqref{eq:sg} and \eqref{eq:s}.

In this paper, we consider the two types of magnetic impurity (Heisenberg type, Ising type) in addition to non-magnetic impurity.
In the case of the isotropic magnetic impurity (Heisenberg type), we replace 
the form factor of the spins in Eq.~\eqref{eq:scrf} as
\begin{align}
  &\delta_{\sg_1 \sg_2} \delta_{\sg_3 \sg_4} \to 
  \bm \sg_{\sg_1\sg_2} \cdot \bm \sg_{\sg_3\sg_4}
  = 
  2\delta_{\sg_1 \sg_4} \delta_{\sg_2 \sg_3} - \delta_{\sg_1 \sg_2} \delta_{\sg_3 \sg_4}
\end{align}
for spin $S = 1/2$.
We can also consider the magnetic impurity with anisotropy in $z$-direction (Ising type):
\begin{align}
    &\delta_{\sg_1 \sg_2} \delta_{\sg_3 \sg_4} \to \sg^z_{\sg_1 \sg_2}\sg^z_{\sg_3 \sg_4}.
\end{align}
Thus, the parameters that control the impurity effect is the scattering strength $\Gamma$ and the type of the impurity potential (non-magnetic or Ising type/Heisenberg type).

In the next section, Sec.~\ref{sec:result}, we will discuss 
the parameter $\Gamma$ dependence of single particle spectra.

\section{Numerical Results for single-particle spectra \label{sec:result}}

In the above, we have formulated the theory of disordered BFS.
Here we explain the method of calculation for the physical quantities such as single-particle excitation spectra.
We also show the numerical results in the following subsections.

\subsection{Single particle spectra near Fermi level \label{sec:spectra}}

From the Green's function, we can calculate the DOS.
Using the retarded Green's function $\check G_{\bm k}(\om + \imu 0^+)$ obtained by the analytic continuation from imaginary axis to real axis, we define the DOS as
\begin{align}
  &D(\om) = -\frac{1}{N \pi} \trace \imag \sum_{\bm k \in \mrm{HBZ}} \check G_{\bm k}(\om + \imu 0^+). \label{eq:dos}
\end{align}
Similar to the Fermi liquid theory of electrons, the low energy contributions are extracted in order to see the detailed structure near the BFSs.
The concrete calculation procedure is shown in Appendix~\ref{sec:calc}.
We write down the final result:
\begin{align}
    D(\om) = D_0 \sum_{C} \int_{C} \diff \bm k A_{\bm k}(\om), \label{eq:ak} 
\end{align}
where
\begin{align}
    &A_{\bm k}(\om) = \sum_{b \in \mrm{BFS}} \frac{V_{\mrm{c}}}{(2\pi)^2 |v_b(\bm k)|D_0} \nt 
    &\times\real \Bigg[\frac{\sgn\imag \ep_{\bm k +}(\om + \imu 0^+) - \sgn\imag \ep_{\bm k -}(\om + \imu 0^+)}{\ep_{\bm k +}(\om + \imu 0^+) - \ep_{\bm k -}(\om + \imu 0^+)} \nt 
    &\times (2\om - [\check \Sigma_{\bm k}'(\om + \imu 0^+)]_{1 1} - [\check \Sigma_{\bm k}'(\om + \imu 0^+)]_{2 2})\Bigg] \label{eq:ak_low}
\end{align}
is the wave-vector-resolved spectral function, and
\begin{align}
    D_0 = \sum_{C} \int_{C} \diff \bm k \frac{V_{\mrm{c}}}{2\pi^2 |v_b(\bm k)|} \label{eq:d0}
\end{align}
is a zero-energy DOS in the clean limit.
$\int_C \diff\bm k$ is the path integral along the BFS, which is taken over the specific path $C = C_\Gamma, C_{\mrm M1}, C_{\mrm M2}$ [See, Fig.~\ref{fig:fermi} (b)].
$\check \Sigma_{\bm k}'$ is the self-energy of bogolon defined by Eq.~\eqref{eq:self_energy_bog} in Appendix~\ref{sec:calc}, where the prime ($'$) symbol indicates the basis of bogolon picture.
$\ep_{\bm k\pm}$ is a quantity dependent on the self-energies defined in Appendix D.
In the clean limit, it reduces to
\begin{align}
    &A^0_{\bm k}
    = \frac{V_{\mrm{c}}}{\pi^2 D_0 |v_b(\bm k)|} \label{eq:ak0}
\end{align}
whose values are plotted in Fig.~\ref{fig:fermi} (c).
Since the $\bm k$ dependence originates only from the velocity $v_b(\bm k)$, $A^0_{\bm k}$ has the four-fold symmetry if the energy eigenvalue $E_{\bm kb}$ is four-fold symmetric in $\bm k$-space.
$A^0_{\bm k}$ in (c) has characteristic peaks in $C_{\mrm M1}$ and $C_{\mrm M2}$ (indicated by green circles).
These peaks correspond to the points where the BFS has large curvature.
In contrast, the behavior of $A^0_{\bm k}$ in $C_{\Gamma}$ is relatively stable, where curve of the BFS is gentle.

\subsection{Born approximation}

Substituting the Green's function in the clean limit to the self-energies in Eqs.~\eqref{eq:sg} and \eqref{eq:s}, we obtain the self-energies in the Born approximation.
We note that the self-energies do not have $\om$-dependence in the present setup (It appears in the self-consistent Born approximation as shown later.).
Inserting the self-energies into Eq.~\eqref{eq:ak_low}, we obtain the spectral function.

\begin{figure}[tb]
    \centering
    \includegraphics[width=8.5cm]{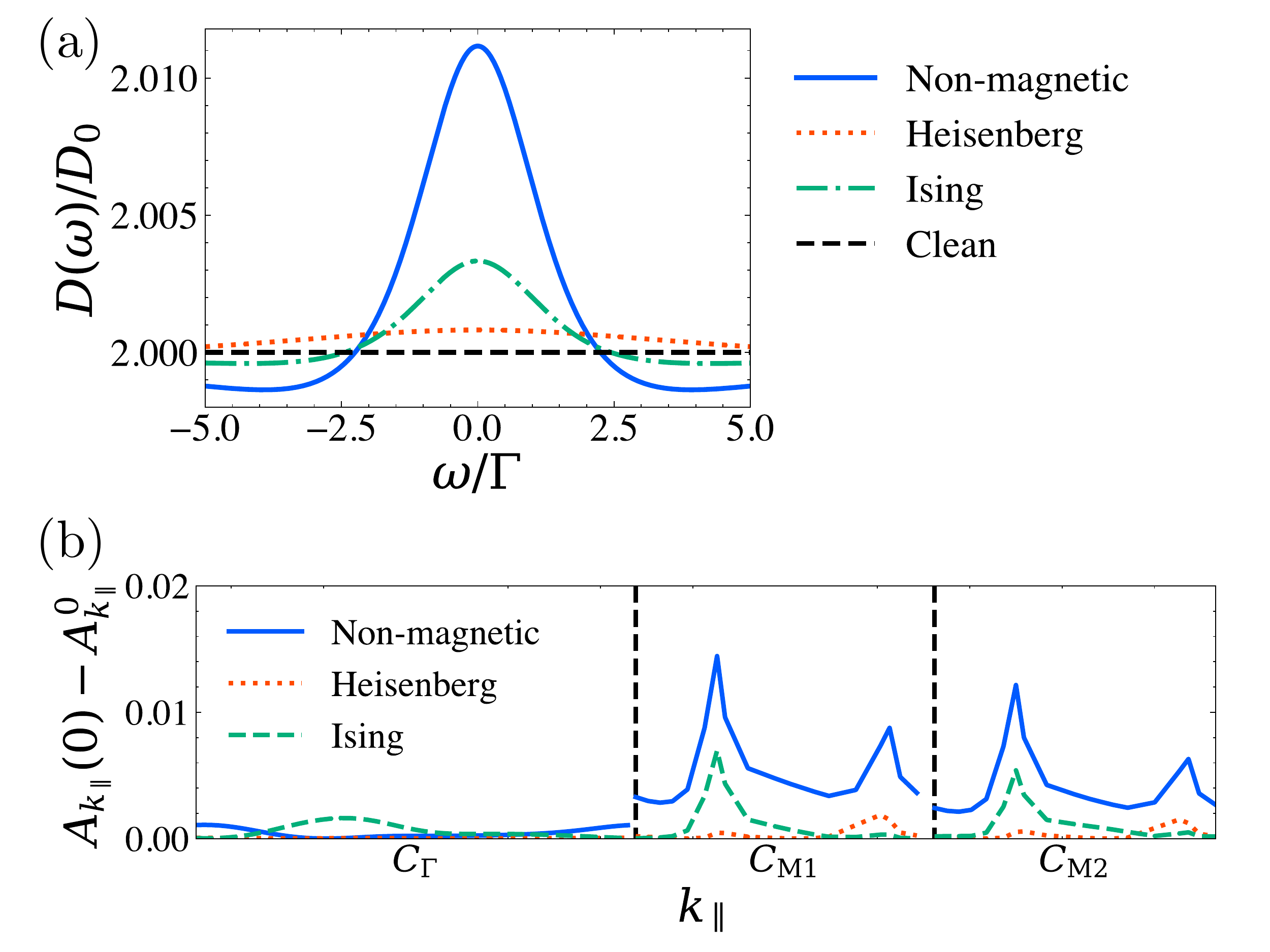}
    \caption{Comparison between different kinds of impurities. (a) Energy dependence of the density of states, (b) Wave-vector dependent spectral function on the BFSs. The horizontal axis of (b) is taken in the same way as Fig.~\ref{fig:fermi} (c).}
    \label{fig:born_mag}
\end{figure}

Figure~\ref{fig:born_mag} (a) shows the DOS defined in Eq.~\eqref{eq:dos} (the more specific form is shown in Appendix~\ref{sec:calc}).
Here the value is normalized by the one in the clean limit Eq.~\eqref{eq:d0}, and we consider the three kinds of impurities explained in Sec.~\ref{sec:imp}-C. 
We find that zero-energy peak in the DOS, which is absent in the clean limit.
It is notable that the peak height does not depend on the  impurity density and the magnitude of the impurity potential,
since $\check \Sigma' \propto \Gamma (\propto n_{\mrm{imp}} U_{\mrm{imp}}^2)$ (See, Eqs.~\eqref{eq:sg}, \eqref{eq:s} and \eqref{eq:uf}.) and then $\Gamma$ is cancelled in Eq.~\eqref{eq:dos_low}.
On the other hand, the peak height changes depending on the type of impurity potential.
With these results, we can roughly estimate the order of this peak height: it is about one percent of the clean-limit DOS $D_0$ in the case of non-magnetic impurity.

Figure~\ref{fig:born_mag} (b) shows the wave-vector resolved spectral functions at each $\bm k$ points on the BFSs.
In order to see the effect of impurity, the difference between the spectral function with and without impurities is plotted.
The differences $A_{k_\parallel}(0) - A_{k_\parallel}^0$ vary depending on the type of impurity, and the relative height vary across different regions within the Brillouin zone.
The results in Fig.~\ref{fig:born_mag} (b) are correlated with Fig.~\ref{fig:fermi} (c): the deviation $A_{k_\parallel}(0) - A_{k_\parallel}^0$ is large when the spectral function in the clean limit is large.

Next we discuss the self-energies near the BFSs.
Since the self-energies of bogolon satisfies the relations $[\check \Sigma'_{\bm k}(\om + \imu 0^+)]_{11} = -[\check \Sigma'_{\bm k}(-\om + \imu 0^+)]_{22}^\ast$ and $[\check \Sigma'_{\bm k}(\om + \imu 0^+)]_{12} = - [\check \Sigma_{\bm k}(-\om + \imu 0^+)]_{21}^\ast$ from Hermiticity and inversion symmetry, the spectral function in Eq.~\eqref{eq:ak_low} is
determine by the following two quantities
\begin{align}
    &-\imu\Gamma_{1\bm k}(\om) = [\check \Sigma_{\bm k}'(\om + \imu 0^+)]_{1 1} \label{eq:bog_1}, \\
    &-\imu\Gamma_{2\bm k}(\om) = [\check \Sigma_{\bm k}'(\om + \imu 0^+)]_{1 2}. \label{eq:bog_2}
\end{align}
$\Gamma_{1\bm k}, \Gamma_{2\bm k}$ are identical respectively to the normal and anomalous self-energies for bogolons at low-energy regime as discussed in Sec.~\ref{sec:overview}.
The zero-energy limit of the spectral function is given as follows
\begin{align}
    &\frac{A_{\bm k}(\om \to 0)}{A_{\bm k}^0} =
    \frac{1}{\sqrt{1 - \big[|\Gamma_{2\bm k}(\om \to 0)| / \real \Gamma_{1\bm k}(\om \to 0)\big]^2}}, \label{eq:dos0}
\end{align}
where the spectral function is controlled by the ratio
$\Gamma_{2\bm k} / \Gamma_{1\bm k}$.
Whereas the $\bm k$ dependence is neglected in Ref.~\cite{Miki21} for simplicity, this paper takes full account of it on the BFSs.

\begin{figure}[tb]
    \centering
    \includegraphics[width=8.5cm]{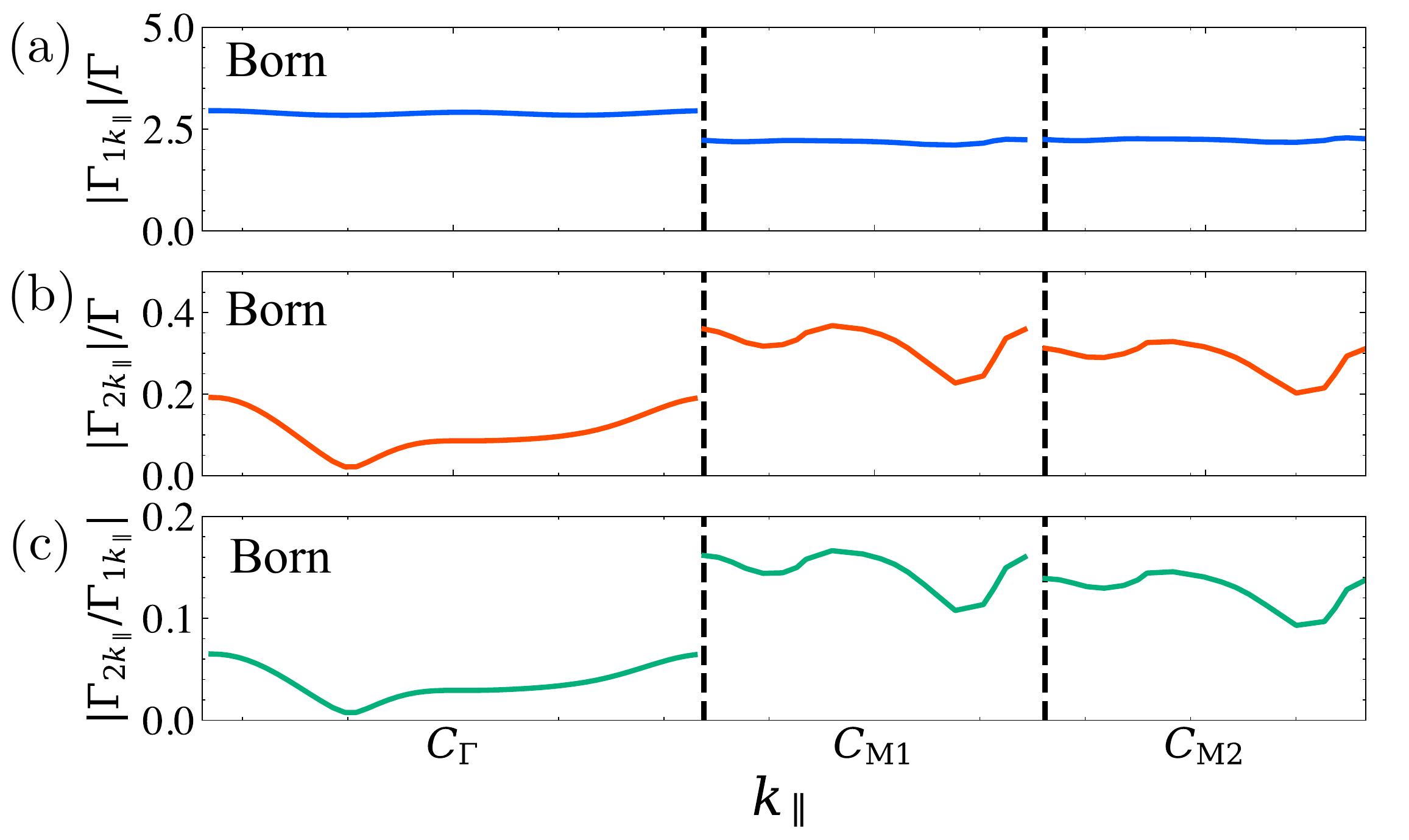}
    \caption{Self-energies of bogolon obtained by the Born approximation for (a) normal part, (b) anomalous part, and (c) the ratio of the normal and anomalous parts. The horizontal axis is taken in the same way as Fig.~\ref{fig:fermi} (c).}
    \label{fig:bog}
\end{figure}

Figure~\ref{fig:bog} shows the wave-vector-dependent self-energies of bogolon for Born approximation, where the $\om$-dependence is absent.
The figures (a), (b), and (c) correspond to the normal part $|\Gamma_{1\bm k}|$, anomalous part $|\Gamma_{2\bm k}|$, and their ratio $|\Gamma_{2\bm k}/\Gamma_{1\bm k}|$, respectively.
The anomalous part $|\Gamma_{2\bm k}|$ has a stronger $k_\parallel$-dependence in comparison to the normal part $|\Gamma_{1\bm k}|$.
Then the $k_\parallel$ dependence of $|\Gamma_{2\bm k}/\Gamma_{1\bm k}|$, which determines the height of the spectral function according to Eq.~\eqref{eq:dos0}, resembles that of $|\Gamma_{2\bm k}|$.
The absolute values can be roughly estimated $|\Gamma_{1\bm k}|/\Gamma \sim 1, |\Gamma_{2\bm k}|/\Gamma \sim 0.1$, and then $|\Gamma_{2\bm k}/\Gamma_{1\bm k}| \sim 0.1$.
Since the square of $|\Gamma_{2\bm k}/\Gamma_{1\bm k}|$ ($\sim 0.01$) determines the spectral functions, the peak value in the DOS becomes 1\% of that in the clean limit as shown in Fig.~\ref{fig:born_mag} (a).

Although the detection of such small change in DOS might be difficult experimentally, our result indicates that, with systematically increasing the impurity scattering, the DOS peak height remains unchanged while the peak width increases.
This behavior originates from the impurity effect characteristic for BFS where the odd-frequency pair potential ($\Gamma_{2\bm k}$) is involved.

We also comment on the relation to our previous work \cite{Miki21}, in which
we studied the impurity effect on BFS by using a simplified low-energy effective model of bogolon.
We have neglected the $\bm k$-dependence of $\Gamma_{1\bm k}$ and $\Gamma_{2\bm k}$ for simplicity and
observed a peak structure in the DOS.
This behavior is qualitatively consistent with the results of the DOS in the present paper.
Here we have further clarified the $k_\parallel$-dependence of the self-energies and estimated the order of magnitude of spectra.

\subsection{Self-consistent Born approximation}

\begin{figure}[tb]
    \centering
    \includegraphics[width=8.5cm]{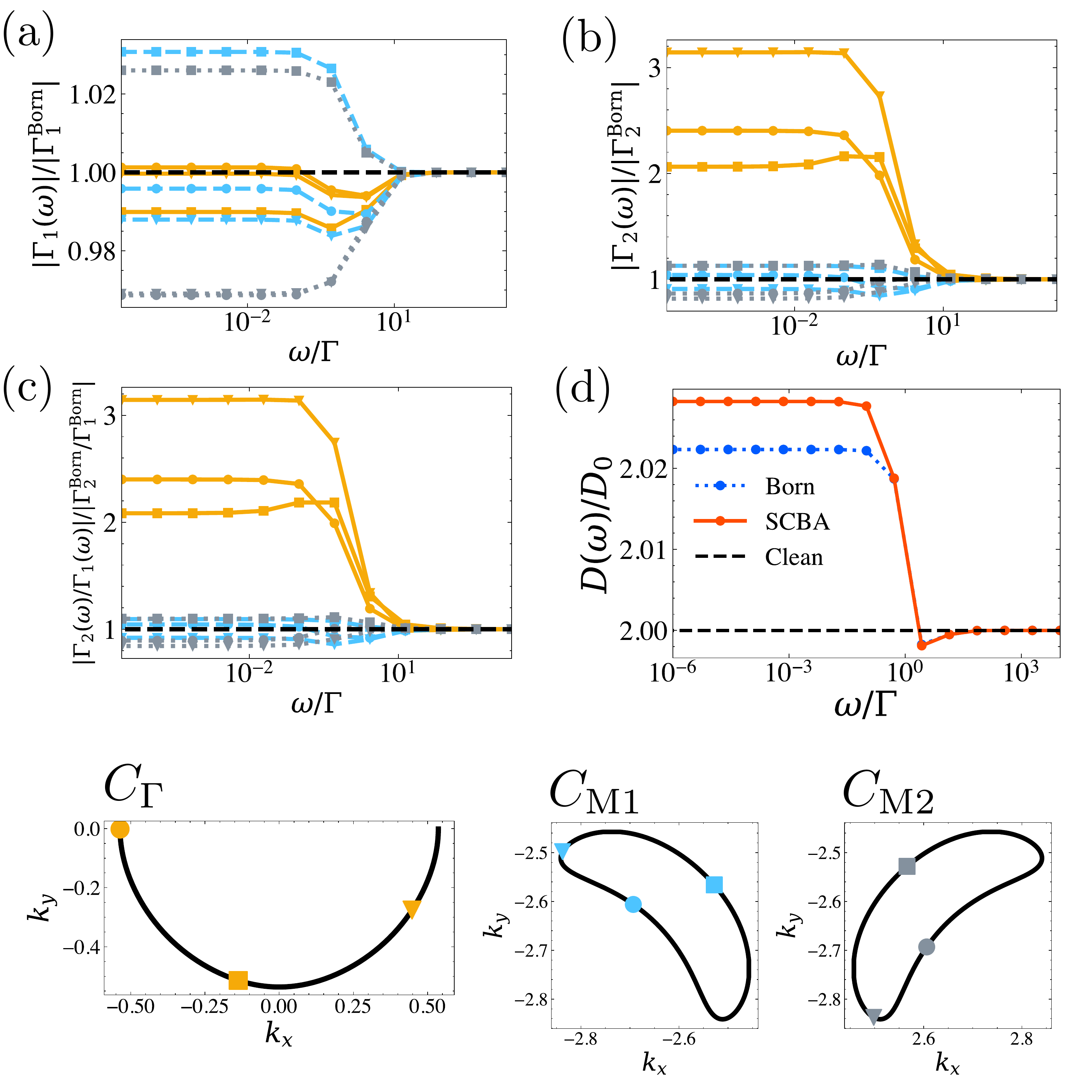}
    \caption{(a)-(c) $\om$-dependence of self-energies for bogolons obtained by the self-consistent Born approximation, which is normalized by the value for Born approximation. 
    The colors and symbols of each line correspond to those of $C_\Gamma, C_{\mrm M1}, C_{\mrm M2}$ shown in the bottom of the figure. 
    (d) Density of states. 
    The number of $k_\parallel$ mesh is $18$ on the BFSs.}
    \label{fig:scba}
\end{figure}

Next we consider the self-consistent Born approximation.
We obtain the self-energies by solving Eqs.~\eqref{eq:sg} and \eqref{eq:s} self-consistently.
The concrete forms of the Green's functions are shown in Appendix~\ref{sec:calc}.

Figures~\ref{fig:scba} (a)-(c) show the $\om$-dependence of the self-energies of bogolon, which are normalized by the value for Born approximation.
We have chosen several $\bm k$ points on the BFSs $C_\Gamma, C_{\mrm M1}, C_{\mrm M2}$ shown in the bottom part of Fig.~\ref{fig:scba}.
We confirm that each quantity coincide to that of Born approximation at large $\om$ as expected.
On the other hand, these quantities change at small frequencies because of the mixing of normal and anomalous parts by the self-consistent calculation.
$|\Gamma_{1\bm k}(\om)|$ is not much changed from Born approximation as seen from Fig.~\ref{fig:scba} (a).
On the contrary, $|\Gamma_{2\bm k}(\om)|$ becomes larger at low frequency for $C_\Gamma$.
Hence $|\Gamma_{2\bm k}(\om)/\Gamma_{1\bm k}(\om)|$ becomes larger.
This behavior results in the larger peak height of the DOS as shown in Fig.~\ref{fig:scba} (d) compared to the Born approximation.
This is because the peak height is determined by $\Gamma_{2\bm k}(\om \to 0)/\Gamma_{1\bm k}(\om \to 0)$ as discussed in the previous subsection.

Finally, we comment on the two kinds of the solutions in self-consistent Born approximation as proposed in Ref.~\cite{Miki21}.
In our former study, we applied the self-consistent Born approximation for an effective low-energy model for bogolons, where $\bm k$ dependence in self-energies is neglected.
With this simplified set up, we have scanned the parameter space and found the two kinds of solutions:
one is a Born-approximation-like solution with $\Gamma_{1,2}(\om) \sim \mrm{Const.}$ (first-kind), and the other is characterized by $\Gamma_{1,2}(\om) \propto 1/\om$ (second-kind) at low $\om$ \cite{Miki21}.
In the present analysis, we obtain the solution of the first-kind as shown in Fig.~\ref{fig:scba} (a)-(c), while the solution of second-kind is not realized at least in our analysis based on iron-based superconductors.

\section{Summary \label{sec:summary}}

In this paper, we have analyzed the impurity effect on Bogoliubov Fermi surface using a realistic model for iron-based materials combined with the Born approximation. We have investigated the detailed structure for the single-particle spectra at low energies. Based on the tight-binding parameters obtained from first-principles calculations, we have calculated the wave-vector-dependent dynamical self-energy focusing on the low-energy regime near the Bogoliubov Fermi surfaces.

We have found that the zero-energy peak appears in the density of states as induced by the off-diagonal self-energy for bogolons. Furthermore, we have estimated an order of magnitude of the peak. 
The peak height is about one percent of the density of states in the clean limit and is independent of the impurity density and the magnitude of the potential in the Born approximation.
On the other hand, the peak height varies with the type of non-magnetic or magnetic impurities. 
These features are unique to Bogoliubov Fermi surface. 
Hence, a systematic study of impurity effects may provide experimental confirmation of the presence of Bogoliubov Fermi surface in $\mrm{Fe(Se,S)}$.
Although we have assumed a specific pair potential in this paper, the present method can be applied to other superconducting states.

\section*{Acknowledgments}

This work was supported by KAKENHI Grants
No. 19H01842, No. 19H05825, and No. 21K03459.

\appendix

\section{Details of the first principles calculations \label{sec:dft}}

We performed the first-principles calculations of $\mrm{FeSe}$ using \texttt{Quantum ESPRESSO} (QE) \cite{Giannozzi17} and constructed the tight-binding Hamiltonian with \texttt{wannier90} \cite{Pizzi20}. 
In QE calculations, we used the exchange-correlation functional proposed by Perdew, Burke, and Ernzerhof \cite{Perdew96}, and the ultrasoft pseudopotentials \cite{Vanderbilt90} provided in \texttt{pslibrary} \cite{DalCorso14}. 
The plane wave cutoff energy and charge density cutoff values were set to be $46\, \mrm{Ry}$ and $240\, \mrm{Ry}$, respectively. 
The crystal structure parameters of $\mrm{FeSe}$ were obtained from experimental data \cite{Bohmer13}. 
However, two $\mrm{Fe}$ sites were placed at $(0,0,0)$ and $(0.5,0.5,0)$ to increase the accuracy of the calculation in \texttt{wannier90}. Then, we constructed maximally localized Wannier functions for ten orbitals of Fe($3d$). 
However, as is well known, the obtained band structure deviates significantly from experimental observations. 
Here, we have adjusted the band structure by referring to Ref.~\cite{Yamakawa16}; The $xz/yz$ and $x^2-y^2$ site energies of the two $\mrm{Fe}$ sites in the unit cell were added by $+0.1 \, \mrm{eV}$ and $+0.04\, \mrm{eV}$, respectively. 
The $(1,0,0)$ and corresponding hopping integrals of the $xz/yz$ and $x^2-y^2$ orbitals were added by $-0.009\, \mrm{eV}$ and $-0.018\, \mrm{eV}$, respectively. 
The Fe(I)-Fe(II) nearest-neighbor hopping integrals for the $xz/yz$ and $x^2-y^2$ orbitals were added by $-0.05\, \mrm{eV}$.

\section{Symmetry operation \label{sec:symmetry}}

We perform symmetry operation for annihilation operator
\begin{align}
    &c_{\gm \sg}(\bm R_{\bm n} + \bm d_a) \nt
    &\to \sum_{\gm_1} (\epn^{\imu\bm \theta \cdot \bm L})_{\gm \gm_1} c_{\gm_1 \sg}(\al(\bm R_{\bm n} + \bm d_a) + \bm b), \label{eq:sym_c}
\end{align}
where $\al$ is a orthogonal matrix (inversion, mirror, rotation, and these combination) with the rotation vector $\bm \theta$, $\bm L$ is a angular momentum for $d$-orbital, and $\bm b$ is a translation vector.
We perform this symmetry operation to Eq.~\eqref{eq:ham_n}, and then obtain
\begin{align}
  &\sum_{\bm n \bm m \gm \gm' \sg} H_{\mrm N \gm \gm'}(\bm R_{\bm n} + \bm d_{a'} - \bm d_a) \nt 
  &\times \sum_{\gm_1 \gm_2} c_{\gm_1 \sg}^\dg(\al(\bm R_{\bm n} + \bm d_a) + \bm b) (\epn^{\imu\bm \theta \cdot \bm L})_{\gm_1 \gm} \nt 
  &\times (\epn^{-\imu\bm \theta \cdot \bm L})_{\gm' , \gm_2} c_{\gm_2 \sg_2}(\al(\bm R_{\bm n} + \bm R_{\bm m} + \bm d_{a'}) + \bm b) \nt
  &= \sum_{\bm n \bm m \gm \gm' \sg} (\epn^{\imu\bm \theta \cdot \bm L})_{\gm_1 \gm} H_{\mrm N \gm \gm'}(\al^{-1}(\bm R_{\bm n} + \bm d_{a'} - \bm b) - \al^{-1}(\bm d_a - \bm b)) \nt 
  &\times (\epn^{-\imu\bm \theta \cdot \bm L})_{\gm' , \gm_2} \sum_{\gm_1 \gm_2} c_{\gm_1 \sg}^\dg(\bm R_{\bm n} + \bm d_a) c_{\gm_2 \sg_2}(\bm R_{\bm n} + \bm R_{\bm m} + \bm d_{a'}).
\end{align}
After the symmetry operation $\al^{-1}(\bm R_{\bm n} + \bm d_{a'} - \bm b)$, we can define new lattice vector $\bm R_{\tilde{\bm n}'}$ and position of sublattice $\bm d_{\tilde{a}'}$, i.e. $\al^{-1}(\bm R_{\bm n} + \bm d_{a'} - \bm b) = \bm R_{\tilde{\bm n}'} + \bm d_{\tilde{a}'}$. Similarly, we write $\al^{-1}(\bm d_{a} - \bm b) = \bm R_{\tilde{\bm n}} + \bm d_{\tilde a}$. Then we obtain
\begin{align}
  &\sum_{\bm n \bm m \gm \gm' \sg} \sum_{\gm_1 \gm_2} (\epn^{-\imu\bm \theta \cdot \bm L})_{\gm_1 \gm} H_{\mrm N \gm \gm'}(\bm R_{\tilde{\bm n}'} - \bm R_{\tilde{\bm n}} + \bm d_{\tilde{a}'} - \bm d_{\tilde a}) \nt
  &\times (\epn^{\imu\bm \theta \cdot \bm L})_{\gm' \gm_2} c_{\gm_1 \sg_1}^\dg(\bm R_{\bm n} + \bm d_a) c_{\gm_2 \sg_2}(\bm R_{\bm n} + \bm R_{\bm m} + \bm d_{a'}).
\end{align}
We can conclude that $H_{\mrm N \gm \gm'}(\bm R_{\bm n} + \bm d_{a'} - \bm d_a)$ needs to have the symmetry
\begin{align}
    &H_{\mrm N \gm \gm'}(\bm R_{\bm n} + \bm d_{a'} - \bm d_{a}) \nt 
    &= 
    \sum_{\gm_1 \gm_2} (\epn^{-\imu\bm \theta \cdot \bm L})_{\gm \gm_1} H_{\mrm N \gm_1 \gm_2}(\bm R_{\tilde{\bm n}'} - \bm R_{\tilde{\bm n}} + \bm d_{\tilde{a}'} - \bm d_{\tilde a})(\epn^{\imu\bm \theta \cdot \bm L})_{\gm_2 \gm'}.
\end{align}
Next, we consider the symmetry operation for the eigenvector $\hat V(\bm k)$.
Using $\hat V(\bm k)$, we perform the unitary transformation from the Wannier basis to the band basis
\begin{align}
    c_{\bm k a\gm\sg} = \sum_{\lam} [\hat V(\bm k)]_{a \gm , \lam} c_{\bm k \lam \sg}, \label{eq:v}
\end{align}
Since the annihilation operator is transformed as Eq.~\eqref{eq:sym_c}, we can write
\begin{align}
    &c_{\bm k \lam \sg}^\dg 
    = \frac{1}{\sqrt{N}}\sum_{a\gm} c_{\gm \sg}^\dg(\bm R_{\bm n} + \bm d_a) \epn^{\imu \bm k \cdot \bm R_{\bm n}} [\hat V(\bm k)]_{a\gm , \lam} \nt
    &\to \frac{1}{\sqrt{N}}\sum_{a\gm\gm_1\bm n} c_{\gm_1 \sg}^\dg(\al(\bm R_{\bm n} + \bm d_a) + \bm b) (\epn^{-\imu \bm \theta \cdot \bm L})_{\gm_1 \gm} \nt 
    &\times \epn^{\imu \bm k \cdot \bm R_{\bm n}} [\hat V(\bm k)]_{a\gm , \lam} \nt 
    &= \frac{1}{\sqrt{N}}\sum_{a\gm\gm_1\bm n} c_{\gm_1 \sg}^\dg(\bm R_{\tilde{\bm n}} + \bm d_{\tilde{a}}) \epn^{\imu (\al \bm k) \cdot \bm R_{\bm \tilde{n}}} (\epn^{-\imu \bm \theta \cdot \bm L})_{\gm_1 \gm} \nt 
    &\times \epn^{\imu \bm k \cdot [\al^{-1}(\bm d_{\tilde{a}} - \bm b) - \bm d_a]} [\hat V(\bm k)]_{\tilde{a}^{-1}\gm , \lam},
\end{align}
where $\bm R_{\tilde{\bm n}} + \bm d_{\tilde{a}'} = \al(\bm R_{\bm n} + \bm d_{a}) + \bm b$ and $\al^{-1}(\bm d_a - \bm b) = \bm R_{\tilde{\bm n}^{-1}} + \bm d_{\tilde{a}^{-1}}$.
Then we can obtain the eigenvector at $\al \bm k$
\begin{align}
    [\hat V(\al \bm k)]_{\tilde{a}\gm , \lam} = \sum_{\gm_1} (\epn^{-\imu \bm \theta \cdot \bm L})_{\gm \gm_1} \epn^{\imu \bm k \cdot [\al^{-1}(\bm d_{\tilde{a}} - \bm b) - \bm d_a]} [\hat V(\bm k)]_{\tilde{a}^{-1}\gm_1 , \lam}. \label{eq:eigenvec_sym}
\end{align}

\section{Cubic harmonics \label{sec:cubic}}

We list the cubic harmonics for $d$-orbital as follows:
\begin{align}
  &\Theta_{{z^2}}(\theta , \varphi) = \sqrt{\frac{5}{16\pi}} (3\cos^2\theta - 1), \\
  &\Theta_{xz}(\theta , \varphi)  = \sqrt{\frac{15}{4\pi}} \sin\theta \cos\theta \cos\varphi, \\
  &\Theta_{yz}(\theta , \varphi)  = \sqrt{\frac{15}{4\pi}} \sin\theta \cos\theta \sin\varphi, \\
  &\Theta_{x^2 - y^2}(\theta , \varphi)  = \sqrt{\frac{15}{16\pi}} \sin^2\theta \cos2\varphi, \\
  &\Theta_{xy}(\theta , \varphi)  = \sqrt{\frac{15}{16\pi}} \sin^2\theta \sin2\varphi.
\end{align}
These functions satisfy the orthogonal relation as
\begin{align}
  \int \diff \theta \diff \varphi\, \sin\theta \Theta_{m}(\theta,\varphi) \Theta_{m'}(\theta,\varphi) = \pi^2\delta_{mm'}.
\end{align}
Furthermore, can evaluate the following integral:
\begin{align}
  &\int \diff \theta \diff \varphi \, \sin\theta \Theta_{\gm_1}(\theta, \varphi) \Theta_{\gm_2}(\theta, \varphi) \Theta_{\gm_3}(\theta, \varphi) \Theta_{\gm_4}(\theta, \varphi) \nt
  &= \frac{5}{28\pi}(\delta_{\gm_1 , \gm_2} \delta_{\gm_3 , \gm_4} + \delta_{\gm_1 , \gm_3} \delta_{\gm_2 , \gm_4} + \delta_{\gm_1 , \gm_4} \delta_{\gm_2 , \gm_3}). \label{eq:cubit_int}
\end{align}
In this basis, angular momentum is given by
\begin{align}
  &L_x = 
  \begin{pmatrix}
    0 & 0 & \sqrt{3}\imu & 0 & 0 \\
    0 & 0 & 0 & 0 & \imu \\
    -\sqrt{3}\imu & 0 & 0 & 0 & 0 \\
    0 & 0 & 0 & 0 & 0 \\
    0 & -\imu & 0 & 0 & 0 \\
  \end{pmatrix}, \\
  &L_y = 
  \begin{pmatrix}
    0 & -\sqrt{3}\imu & 0 & 0 & 0 \\
    \sqrt{3}\imu & 0 & 0 & -\imu & 0 \\
    0 & 0 & 0 & 0 & -\imu \\
    0 & \imu & 0 & 0 & 0 \\
    0 & 0 & \imu & 0 & 0 \\
  \end{pmatrix}, \\
  &L_z = \begin{pmatrix}
    0 & 0 & 0 & 0 & 0 \\
    0 & 0 & -\imu & 0 & 0 \\
    0 & \imu & 0 & 0 & 0 \\
    0 & 0 & 0 & 0 & -2\imu \\
    0 & 0 & 0 & 2\imu & 0 \\
  \end{pmatrix}.
\end{align}

\section{Calculation of single particle spectra \label{sec:calc}}

\subsection{Density of state \label{sec:calc_dos}}

We perform the summation of $\bm k$ in Eq.~\eqref{eq:dos} focusing on the low energy.
It is convenient to introduce the short-hand notation for the self-energy of bogolon as
\begin{align}
    \check \Sigma_{\bm k}'(\imu \om_n) = \check U(\bm k)^\dg \check \Sigma_{\bm k}(\imu \om_n) \check U(\bm k). \label{eq:self_energy_bog}
\end{align}
Using this, the Green's function is written as follows
\begin{align}
    \check G_{\bm k}(\imu\om_n) &= \check U(\bm k) \Bigg[\imu\om_n\check 1 - \check{\mcal{E}}(\bm k) - \check \Sigma_{\bm k}'(\imu\om_n) \Bigg]^{-1} \check U(\bm k)^\dg. \label{eq:green_u}
\end{align}
Since the contribution near the BFSs becomes larger, we extract two low-energy bands of bogolon $b \in \mrm{BFS} = 1, 2$ which make the BFSs.
For the calculation of the DOS, we change the coordinate of $\bm k$ as $(k_x , k_y) \to  (k_\parallel , k_\perp)$, where $k_\parallel$ is a parallel component to the BFSs and $k_\perp$ is its perpendicular component.
Then the integral around the path $C$ can be rewritten by using the energy of bogolon $\ep\, (= E_b)$ as $\diff k_{\be\parallel} \diff k_{\be\perp} = \diff k_{\be\parallel} \diff \ep/|v_b(k_{\be\parallel})|$ with Fermi velocity of bogolon $v_b(k_{\be\parallel})$.
A similar method is used in Ref.~\cite{Tamura20}.
There are three paths of the BFSs $C = C_{\Gamma}, C_{\mrm M1}, C_{\mrm M2}$ [See, Fig.~\ref{fig:fermi} (a) and (b).] in the half-Brillouin zone.

With these preliminaries, we finally obtain
\begin{widetext}
\begin{align}
  &D(\om) = -\frac{1}{\pi} \imag \sum_C \int_{C} \diff \bm k \frac{V_{\mrm{c}}}{2\pi^2 |v_b(\bm k)|}
  \sum_{b'\in \mrm{BFS}} \int_{-\om_c}^{\om_c} \diff\ep \Bigg[(\om + \imu0^+)\check 1 - \check{\mathcal{E}}(\bm k)
  - \check \Sigma_{\bm k}'(\om + \imu 0^+) \Bigg]^{-1}_{b' b'}, \label{eq:dos_t}
\end{align}
where $\om_c$ is a cut-off energy and $V_c = V/N$.
$\int_C \diff\bm k$ is the integral of $k_\parallel$ direction, which is taken over the path $C$.
Below we evaluate a summation of $b'$ in Eq.~\eqref{eq:dos_t}.
The $2\times 2$ matrix which enclosed in $[\cdots]$ is expressed as
\begin{align}
  &\Bigg[(\om + \imu0^+)\check 1 - \check{\mathcal{E}}(\bm k)
  - \check \Sigma_{\bm k}'(\om + \imu 0^+) \Bigg]^{-1}_{b' b'} \nt
  &=-\frac{1}{(\ep - \ep_{\bm k +}(\om + \imu 0^+))(\ep - \ep_{\bm k -}(\om + \imu 0^+))}
  \begin{pmatrix}
    \om + \imu 0^+ + \ep - [\check \Sigma_{\bm k}'(\om + \imu 0^+)]_{2 2} & [\check \Sigma_{\bm k}'(\om + \imu 0^+)]_{1 2} \\
    [\check \Sigma_{\bm k}'(\om + \imu 0^+)]_{2 1} & \om + \imu 0^+ - \ep - [\check \Sigma_{\bm k}'(\om + \imu 0^+)]_{1 1}
  \end{pmatrix}_{bb'}, \label{eq:g_mat}
\end{align}
where we use $\ep_1 = -\ep_2 \equiv \ep$ for the inversion symmetry, and $\ep_{\bm k \pm}$ is defined by
\begin{align}
    &\ep_{\bm k \pm}(z) = \frac{1}{2}\Big(-[\check \Sigma_{\bm k}'(z)]_{1 1} + [\check \Sigma_{\bm k}'(z)]_{2 2} \Big)
    \pm \dfrac{1}{2}\Bigg[\left([\check \Sigma_{\bm k}'(z)]_{1 1} - [\check \Sigma_{\bm k}'(z)]_{2 2}\right)^2 \nt
    &+ 4\Big((\om + \imu 0^+ - [\check \Sigma_{\bm k}'(z)]_{1 1})(\om + \imu 0^+ - [\check \Sigma_{\bm k}'(z)]_{2 2}) - [\check \Sigma_{\bm k}'(z)]_{1 2} [\check \Sigma_{\bm k}'(z)]_{2 1}\Big)\Bigg]^{1/2}.
\end{align}
Performing the integration of $\ep$ in Eq.~\eqref{eq:dos_t}, we obtain
\begin{align}
    &D(\om) = D_0 \sum_{C} \int_{C} \diff \bm k \frac{V_{\mrm{c}}}{2\pi^2 |v_b(\bm k)|D_0} 
    \real \Bigg[\frac{\sgn\imag \ep_{\bm k +}(\om + \imu 0^+) - \sgn\imag \ep_{\bm k -}(\om + \imu 0^+)}{\ep_{\bm k +}(\om + \imu 0^+) - \ep_{\bm k -}(\om + \imu 0^+)} \nt
    &\times (2\om - [\check \Sigma_{\bm k}'(\om + \imu 0^+)]_{b_1 b_1} - [\check \Sigma_{\bm k}'(\om + \imu 0^+)]_{b_2 b_2})\Bigg]. \label{eq:dos_low}
\end{align}

\subsection{Self-energies}

We proceed to evaluation of the self-energies.
We deal with the $\bm k$ summation in a similar manner to Eq.~\eqref{eq:dos_low}. 
Then we rewrite the self-energy Eq.~\eqref{eq:sg} as
\begin{align}
  &\Sigma_{\bm k \lam \sg , \lam' \sg'}(\om + \imu 0^+)
  \simeq \Gamma D_0\sum_C \int_C \diff\bm k_1 \frac{V_{\mrm{c}}}{2\pi^2 |v_b(\bm k_1)|D_0} \nt
  &\times\Bigg(\mathscr{F}_{\lam \sg, \lam_1 \sg_1, \lam_2 \sg_2, \lam' \sg'}(\bm k , \bm k_1, \bm k_1, \bm k) \sum_{b' , b''} [\check U(\bm k_1)]_{\lam \sg,b'} [\check U(\bm k_1)^\dg]_{b'' ,\lam' \sg'}
  \int_{-\om_c}^{\om_c} \diff\ep \left[(\om + \imu0^+)\check 1 - \check{\mathcal{E}}(\bm k_1) - \check \Sigma_{\bm k_1}'(\om + \imu 0^+) \right]^{-1}_{b' b''} \nt
  &-\mathscr{F}_{\lam \sg, \lam_1 \sg_1, \lam_2 \sg_2, \lam' \sg'}(\bm k , -\bm k_1, -\bm k_1, \bm k) \sum_{b' , b''} [\check U(\bm k_1)]_{\lam \sg + M,b'} [\check U(\bm k_1)^\dg]_{b'' ,\lam' \sg' + M}
  \int_{-\om_c}^{\om_c} \diff\ep \left[(-\om - \imu0^+)\check 1 - \check{\mathcal{E}}(\bm k_1) - \check \Sigma_{\bm k_1}'(-\om - \imu 0^+) \right]^{-1}_{b' b''}
  \Bigg).
\end{align}
We extract the contribution near the BFSs in a similar manner to Eq.~\eqref{eq:g_mat}.
Then, we obtain
\begin{align}
  &\Sigma_{\bm k \lam \sg , \lam' \sg'}(\om + \imu 0^+)
  = -\imu\pi \Gamma \sum_{\lam_1 \lam_2 \sg_1 \sg_2} \sum_{C} \int_C \diff \bm k_1\, 
  \frac{V_{\mrm{c}}}{2\pi^2 |v_b(k_{1\be\parallel})|D_0} \nt &\times\Bigg(\mathscr{F}_{\lam \sg, \lam_1 \sg_1, \lam_2 \sg_2, \lam' \sg'}(\bm k , \bm k_1, \bm k_1, \bm k)
  \sum_{b'  b''} [\check U(\bm k_1)]_{\lam_1 \sg_1 , b'} [G_{\bm k_1}^{\mrm{bog}}(\om + \imu 0^+)]_{b' b''} [\check U(\bm k_1)^\dg]_{b'' , \lam_2 \sg_2} \nt
  &- \mathscr{F}_{\lam \sg, \lam_1 \sg_1, \lam_2 \sg_2, \lam' \sg'}(\bm k , -\bm k_1, -\bm k_1, \bm k)
  \sum_{b'  b''} [\check U(\bm k_1)]_{\lam_2 \sg_2 + M , b'} [G_{\bm k_1}^{\mrm{bog}}(-\om - \imu 0^+)]_{b' b''} [\check U(\bm k_1)^\dg]_{b'' , \lam_1 \sg_1 + M} \Bigg),
  \label{eq:scba}
\end{align}
where 
\begin{align}
    &G_{\bm k}^{\mrm{bog}}(z) = \sgn\imag \ep_{\bm k -}(z) \sg^z + \frac{\sgn\imag \ep_{\bm k +}(z) - \sgn\imag \ep_{\bm k -}(z)}{\ep_{\bm k +}(z) - \ep_{\bm k -}(z)} 
    \begin{pmatrix}
    z - [\check \Sigma_{\bm k}'(z)]_{2 2} + \ep_{\bm k +}(z) & [\check \Sigma_{\bm k}'(z)]_{1 2} \\
    [\check \Sigma_{\bm k}'(z)]_{2 1} & z - [\check \Sigma_{\bm k_1}'(z)]_{1 1} - \ep_{\bm k +}(z)
    \end{pmatrix}.
\end{align}
Solving Eq.~\eqref{eq:scba} on the BFSs, we can determine $\Sigma_{\bm k \lam \sg , \lam' \sg'}(\om + \imu 0^+)$ self-consistently.
We also calculate $S_{\bm k}, \bar \Sigma_{\bm k}, S_{\bm k}^\dg$ in a similar manner.
Inserting these self-energies into Eq.~\eqref{eq:ak}, we obtain the spectral function $A_{\bm k}(\om)$.

\end{widetext}

\end{document}